\input ./style/arxiv-general.cfg
\documentclass[labelcite,MSNbibl,nameyear,dvips]{arxstspdf}
\makeatletter
   \@ifpackageloaded{graphicx}{}{\usepackage{graphicx}}
\makeatother
\usepackage{url,breakurl}
\usepackage{flushend}
\usepackage{stfloats}
\volume{30}
\issue{2}
\pubyear{2015}
\firstpage{268}
\lastpage{295}
\doi{10.1214/14-STS509}% Updated by VTEXPTS2LaTeX.exe, 20.01.2015 15:00
\docsubty{FLA}


\begin{document}
\begin{frontmatter}

\title{A Conversation with Jerry Friedman}%\thanksref{T1}
% kai straipsnis turi susijusiu diskusiju ir rejoinder'iu
%\relateddois{T1}{Discussed in \relateddoi{d}{10.1214/00-STSXXX} ...; rejoinder at \relateddoi{r}{10.1214/00-STSXXXX}.}
\runtitle{A conversation with Jerry Friedman}
%\pdftitle{}

\begin{aug}
% Corresponding author: Nicholas Fisher - Nicholas.Fisher@sydney.edu.au% Updated by VTEXPTS2LaTeX.exe, 20.01.2015 13:48
\author[A]{\fnms{N. I.}~\snm{Fisher}\corref{}\ead[label=e1]{Nicholas.Fisher@sydney.edu.au}}%,
%\author[]{\fnms{}~\snm{}\ead[label=]{}}
% \and
%\author[]{\fnms{}~\snm{}\ead[label=]{}}
\runauthor{N. I. Fisher}
%\pdfauthor{}

\affiliation{University of Sydney}

\address[A]{Nicholas Fisher is Visiting Professor of Statistics,
School of Mathematics and Statistics F07,
University of Sydney, NSW 2006, Australia \printead{e1}.}\vspace*{-6pt}
%\address[]{ \printead{}.}
\end{aug}

% ABSTRACT
\begin{abstract}
Jerome H. Friedman was born in Yreka, California, USA, on
December 29, 1939. He received his high school education at Yreka High
School, then spent two years at Chico State College before transferring to
the University of California at Berkeley in 1959. He completed an
undergraduate degree in physics in 1962 and a Ph.D. in high-energy particle
physics in 1968 and was a post-doctoral research physicist at the Lawrence
Berkeley Laboratory during 1968--1972. In 1972, he moved to Stanford Linear
Accelerator Center (SLAC) as head of the Computation Research Group,
retaining this position until 2006. In 1981, he was appointed half time as
Professor in the Department of Statistics, Stanford University, remaining
half time with his SLAC appointment. He has held visiting appointments at
CSIRO in Sydney, CERN and the Department of Statistics at Berkeley, and has
had a very active career as a commercial consultant. Jerry became Professor
Emeritus in the Department of Statistics in 2007. Apart from some 30
publications in high-energy physics early in his career, Jerry has
published over 70 research articles and books in statistics and computer
science, including co-authoring the pioneering books \textit{Classification
and Regression Trees} and \textit{The Elements of Statistical Learning}.
Many of his publications have hundreds if not thousands of citations
(e.g., the CART book has over 21,000). Much of his software is
incorporated in commercial products, including at least one popular search
engine. Many of his methods and algorithms are essential inclusions in
modern statistical and data mining packages. Honors include the following:
the Rietz Lecture (1999) and the Wald Lectures (2009); election to the
American Academy of Arts and Sciences (2005) and the US National Academy of
Sciences (2010); a Fellow of the American Statistical Association; Paper of
the Year (\textit{JASA} 1980, 1985; \textit{Technometrics} 1998, 1992); Statistician of the
Year (ASA, Chicago Chapter, 1999); ACM Data Mining Lifetime Innovation
Award (2002), Emanuel \& Carol Parzen Award for Statistical Innovation
(2004); Noether Senior Lecturer (American Statistical Association, 2010);
and the IEEE Computer Society Data Mining Research Contribution Award
(2012).

The interview was recorded at his home in Palo Alto, California during 3--4 August 2012.\vspace*{-6pt}
\end{abstract}

% KEYWORDS
\end{frontmatter}

%s1 #&#
\section{Early Days (1939--1959)}

\label{sec1}

\textbf{NF:} Welcome Jerry. Let's begin at the beginning, which was
not in this part of California.

\textbf{JF:} That's correct. I grew up in a tiny town near the Oregon
border called Yreka: it's ``bakery'' spelled backwards without the ``b.''
Yreka Bakery is a palindrome\ldots  and there \textit{was} a Yreka Bakery in
Yreka.

\textbf{NF:} What were your parents doing?

%f1 #&#
\begin{figure}

\includegraphics{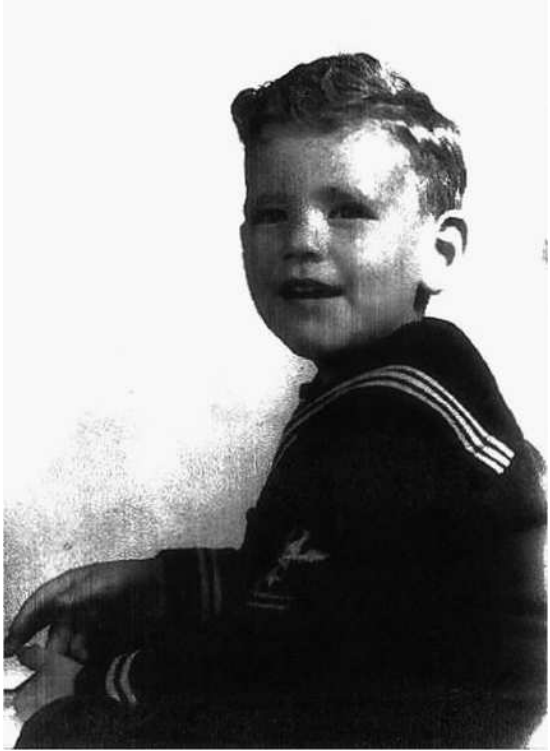}

\caption{Early days---Yreka.}%\label{fig1}
\end{figure}

\textbf{JF:} My mother was a housewife and my father, along with his
brother, owned a laundry and dry-cleaning establishment there that they and
my grandparents founded in the 1930s.

\textbf{NF:} Were your grandparents born in America?

\textbf{JF:} No, one set was born in the Ukraine I think; I'm not sure
where the other set was born. They certainly weren't born in the US, as
they all had heavy accents.

\textbf{NF:} Do you have siblings?

\textbf{JF:} One brother slightly younger than me. He's now retired and
living in LA. He was an accountant for most of his life.

\textbf{NF:} How was school?

\textbf{JF:} School was okay. I was a dramatic underachiever. I wasn't very
interested in school; I was mainly interested in electronics, so I was into
amateur radio, building radio electronics---transmitters, receivers and
that kind of thing---as a kid. This was very unusual for someone in
Yreka. I was really an outlier, but I thought electronics was fascinating,
to be able to talk with people on the other side of the world with no
wires. Now, it's just taken for granted. In those days short wave radio was
the only way to do it. When I was really young in grammar school---10 to
13---I used to build crystal sets all the time. Then I graduated to vacuum
tubes, transmitters and receivers. It's very different electronics than
today. Vacuum tubes operate at very high voltage. So often while you're
poking around trying to see why a circuit isn't working, all of sudden you
pick yourself up on the other side of the room because you touched a place
at about 400 or 500 volts. Today's electronics run at 5 volts. I remember
bugging the math teacher in middle school to teach me square roots because
I needed that to understand some things in this electronics book that I was
reading.

\textbf{NF:} Did you have anybody that you could talk to about this
stuff?

\textbf{JF:} Yes, I had a friend whose father was in amateur radio and knew
a lot about electronics, so I could talk with him about it.

My father went to talk to the principal before I graduated high school and
asked what he should do with me. The principal said, ``Well he's not going
to make it in college. You might try Chico State and when he flunks out you
can put him in the army.'' So that is how I got to Chico State. Its claim
to fame now is that it's where Sierra Nevada Pale Ale beer is brewed.

\textbf{NF:} What was your view of this opinion?

\textbf{JF:} I didn't want to go to Chico State, I wanted to go to
Berkeley. So we struck an agreement that I would go to Chico for two years
and if I wasn't doing too badly, I could consider transferring to Berkeley.
My father was right about that. He wasn't often right, but he was right
about that. At that time Chico State was one of our country's biggest and
best known party schools, not big in size, but its reputation as a party
school was well deserved. There were big parties every night. I~used to
look forward to summer vacations when I could get relief from all those
parties. Every night we drank an enormous amount. There were no drugs
around at that time, but there was lots of alcohol. When I left for
Berkeley two years later I was ready to do something more serious, which
might not have happened if I had gone directly to Berkeley.

\textbf{NF:} Did you actually go to Chico State wanting to learn
something specific?

\textbf{JF:} I wasn't sure what I wanted to be, either a chemist or an engineer. I think
I wanted to be a chemist and I took the elementary chemistry course. I
remember that we were learning how to test for acidity using litmus paper,
which is a real ordeal, and I noticed that the engineering students were
taking the same lectures as us and they were in the same lab, but their lab
was not as intense as ours and they were using some sort of\vadjust{\goodbreak} meter. You put
the meter in the solution and it displayed the pH. I said, ``I like
that,''
so I switched to engineering and actually did engineering at Chico. But
there was a very, very good physics professor there who got me very
interested in physics, so when I transferred to Berkeley I decided to study
physics.

%s2 #&#
\section{UC Berkeley 1959--1972}
\label{sec2}
\textbf{NF:} You then spent the next two years as an undergraduate
at Berkeley. How well did you do?

\textbf{JF:} I think it actually took me two and a half years. I~was
working my way through school. I had no money. I did fairly well. Those
were the days before grade inflation, so I had about a $\mathrm{B}{+} / \mathrm{A}-$
average, which in those days was considered good. Now they get very
impatient if you don't have straight A's, but in those days A's weren't as
easy to get. (See the anecdote \textit{Undergraduate days at Berkeley in
the early `sixties} in \cite*{9}.)

\textbf{NF:} Let's move on to your transition from undergraduate to
graduate student. You're at the end of your undergraduate program and
you`re now deciding what to do. What was your passion of the day?

\textbf{JF:} I wanted to go into Physics. I thought it very interesting and
I couldn't find anything else I found more interesting. I never took a
statistics course.

\textbf{NF:} There was no doubt that you wanted to do it at
Berkeley?

\textbf{JF:} Yes, I loved Berkeley, still do. I like being in the Bay Area.
However, there was a problem.

In those days there was the military draft. Since I had taken an extra
semester to go through undergraduate school, I was ineligible for an
automatic deferment through graduate school. And you had to be in school to
avoid being drafted into the Army. I thought graduate school infinitely
preferable to the army. So for a while I was worried that I would be
drafted because I was classified 1A, healthy and ready to go. I even went
down to the Oakland Induction Center and had my pre-induction physical and
so I figured: this is it, I'm going into the army. Vietnam wasn't big then,
so that was not an issue I was worried about. Learning physics seemed to be
more fun than the army would be. One day I received my new draft
card---they reissued them every year or something like that---and instead of
saying 1A, it said 2E, which meant student deferment. So I had a dilemma
because I thought maybe it was a typographical error. The next time I was
in Yreka, I was torn between either keeping my mouth shut and hoping they
wouldn't discover the mistake, or going up to the Draft Board and asking
them if it was real. I finally decided I'd better find out. The secretary
of the Draft Board said, ``You are 2E,'' and when I looked at her puzzled,
she said, `Well, the Draft Board decided that since you worked your way
through school, it's okay that you took an extra semester to get
through.''

\textbf{NF:} Virtue is more than its own reward.

\textbf{JF:} I guess so. Also, they're given quotas to fill. There are many
kids in Yreka who don't go to college. In fact, in those days there were
very few, so there were lots of young men not in college whom they could
induct. They didn't necessarily need me to fill their quota.

\textbf{NF:} Was it hard to get into graduate school?

\textbf{JF:} I don't know, I think it was, but I wasn't very responsible.
Berkeley physics was the only graduate department I applied to. You should
apply everywhere, but it was the only one I applied to. If I hadn't been
accepted, I would have gone into the army.

\textbf{NF:} What was the view of your parents about pursuing
graduate studies rather than going back and helping out in the business?

\textbf{JF:} Oh, I really knew I wasn't going back to Yreka. Mack Davis,
who is a country singer/songwriter, grew up in Lubbock, Texas. He was once
asked what it was like to grow up in Lubbock. He said, ``Well, happiness is
Lubbock in your rear view mirror,'' and that's the way I usually thought
about Yreka. It was a nice place and all, but it wasn't the place for me.

\textbf{NF:} How did your Ph.D. studies go?

\textbf{JF:} They went well. As things got more difficult my grade point
average seemed to go up rather than down and I really enjoyed it; I loved
doing it. I worked harder and of course there was always the military draft
there if you flunked out. The deferment was good as long as you were in
school. Fortunately for me, I didn't flunk out and I really enjoyed
learning physics.

During the summers I'd worked at radio stations, but in the winter when I
was at school I worked in the library stacking books, which I didn't really
like that much. My roommate mentioned that there were these great jobs at
the Lawrence Berkeley Radiation Laboratory. They did manual pattern
recognition on bubble chamber images of elementary particle reactions. They
needed people to scan the film and pick out the particular patterns that
they were looking for. It was a great job, a bit boring, but it paid much
better than the library, and so I went up there. That's when I started
getting interested in high-energy physics. The leader of the group was
Louis Alvarez. At the time Alvarez hadn't yet received his Nobel Prize. He
received it later in 1968 when I was a graduate student in his group. After
I got my degree, he and his son were the ones who came up with the meteor /
dinosaur extinction theory. One of the smartest men I've ever met.

\textbf{NF:} Did you end up working with him?

\textbf{JF:} No, I worked with Ron Ross, one of the professors in his
group. I worked there as a bubble chamber scanner for a while. Then when I
had to choose a thesis topic there were two reasons for going into
high-energy physics. One was the Alvarez Group. The other one was that in
the courses that I took in the first two years my weakest subject was
quantum mechanics. I~thought if I went into high energy-particle physics, I
would really have to learn quantum mechanics well.

\textbf{NF:} Were you doing any computing at this stage?

\textbf{JF:} I didn't do any computing\ldots  well, actually I did, around
1962. The way I started computing is an interesting story. I was there as a
scanner and one of the more advanced physics graduate students would
sometimes ask me to do little tasks for him besides the scanning. One time
he asked me to draw a scatter plot. He gave me a piece of graph paper, a
pen and a list of the pairs of numbers. He said, ``What you do is for each
pair of numbers, find the corresponding point on the graph and you put a
dot there with the pen.'' I was doing this for a while and of course I'd
repeatedly mess up and have to start over again. One of the other students
said, ``You know, down on the first floor they have a thing called a
computer and it has a cathode ray tube hooked up to it, and it
automatically makes scatter plots. You can write a program to place the
points on the cathode ray tube. A camera then photographs the tube so you
can take a slide of this scatter plot and print it.'' I thought,
\textit{Boy, is that a good idea!} I got a book about programming computers
and I drew my scatter plots with ease.

\textbf{NF:} What were you programming in?

\textbf{JF:} Machine language and Fortran. Fortran was brand new then and
the only high-level programming language. It was very controversial because
real programmers didn't program in Fortran, they programmed in machine
(assembly) language. There was a sign over the entrance to the programming
group office that said ``Any program that can be written in Fortran
deserves to be.'' I guess that's still true today.

\textbf{NF:} What was the nature of the hardware?

\textbf{JF:} The first computer that I actually programmed was a vacuum
tube computer (it wasn't even a discrete transistor computer) called an IBM
704. It had magnetic core memory. There was also an IBM 650 with rotating
drum memory. I liked the 650, even though it was much slower, because for
that you could just walk up and use it. With the 704 you had to book time
and wait to get your job run. The whole thing at Berkeley used punch cards.
I didn't see a text editor until I went to SLAC.

The greatest invention I ever saw was the terminal with the backspace key.
With punch cards, if you make a mistake, you've got to throw the card away
and start over again from the beginning. In the Alvarez Group I was one of
those who did most of the programming. In those days, it was considered
sissy work to some extent. Real physicists built hardware---detectors,
particle beams, etc. Programming was sissy work. High-energy
physicists don't think that way any more because most of them do
programming. But I liked programming much better than building hardware.

\textbf{NF:} What were you doing in your Ph.D. studies?

\textbf{JF:} It was part of a large physics experiment in the 72-inch
hydrogen bubble chamber, which was the same detector that produced the film
I was scanning before. I studied a particular reaction for my thesis:
reactions involving the $k^{-}$ meson.

\textbf{NF:} What sort of hard skills was this calling on,
mathematical skills, computational skills?

\textbf{JF:} Certainly computational skills and understanding the
theoretical physics of the time, which did involve some math. You had to
build a program, and that meant figuring out the algorithms to write the
program. While I was there as a graduate student I wrote a suite of
exploratory data analysis programs that almost everyone in high-energy
physics was using.

\textbf{NF:} So you were actually writing a statistical package.

\textbf{JF:} Yes. Physicists didn't do much hypothesis testing and things
like that; it was mostly exploratory, automatically making scatter plots,
histograms, various other kinds of displays mostly displayed on hardware of
the time, which was mostly this line printer output. \textit{Kiowa} (that's
the name of an Indian tribe) was a package that I wrote. It was the
standard statistical package in high-energy physics all over the world, for
many years. I also wrote a fast general-purpose Monte Carlo program called
\textit{Sage}. Physicists did a lot of Monte Carlo for simulating particle
reactions. I was still getting enquiries about \textit{Sage} twenty years
later, and I believe that some people are still using it.

\textbf{NF:} At some point during your computing activities you
came across Maximum Likelihood.

\textbf{JF:} That's probably when I first really started getting interested
in statistics. There was a physicist, Frank Solmitz, in the Alvarez group
who knew a lot about statistics. He'd written a little technical report
about fundamental statistics for physicists and I thought that was really
interesting. Then another guy, Jay Orear, who was also a physicist, wrote a
little note on maximum likelihood model fitting (\cite{33}). We were
fitting a lot of models and he knew about least squares. I thought that
maximum likelihood was the most elegant idea I had ever seen and it sort of
perked my interest in statistics. Of course it was invented by Fisher, but
I didn't know that; I thought that Jay Orear invented it.

\textbf{NF:} When did you graduate?

\textbf{JF:} I got my degree in 1968 and then they considered me a good
graduate student, so they wanted to hire me as a postdoc physicist at
Berkeley. Postdocs in those days could run forever and they did for a lot
of people. So I stayed until 1972 in the same Alvarez group doing much the
same kind of things, different experiments but basically the same stuff. By
then SLAC (Stanford Linear Accelerator Center) had come online and so I was
involved in an experiment that was running at SLAC while I was at
Berkeley.

\textbf{NF:} Had you started interacting with SLAC?

\textbf{JF:} Well, not really, I mean the data was taken to SLAC, but I
never really went down to SLAC much except to watch the beam. Watching the
beam means that you are taking data; it's a beam of electrons (at Berkeley
it was a beam of protons) and it smashes into matter and then the reaction
products come out and they're detected by particle detectors. There's a
huge amount of electronics controlling all that. So someone has to be in
the control room monitoring the electronics to be sure that everything is
okay and that you're still taking the data at a reasonable rate.

\textbf{NF:} When had SLAC been set up?

\textbf{JF:} SLAC had been built in the sixties, it may have started in the
fifties and it came online in the mid-sixties (1966). This was one of the
first experiments at SLAC. It was an electron machine, so we were in
collaboration with some SLAC people at Berkeley. Our bubble chamber was
moved to SLAC. The data was taken there and brought to Berkeley to be
scanned, measured and analyzed. I didn't spend much time at SLAC during
that period.

\textbf{NF:} Why were the data going to Berkeley?

\textbf{JF:} Because that is the way high-energy physics works even today.
There is a lot of data to analyze, it is very labor-intensive, and so you
spread the work around and it gets done faster.

\textbf{NF:} In other words, distributed computing?

\textbf{JF:} In a sense, yes. Also, these experiments were very expensive
to run, so people like to get together and do it in collaboration. In those
days there were collaborations of tens of physicists, now there are
collaborations of dozens of laboratories.

%s3 #&#
\section{The Move to SLAC (1972)}

\textbf{NF:} Why did you move to SLAC?

\textbf{JF:} Well, we had a new director of the Research Division at
Berkeley who decided that postdocs should not stay on forever and that
three years was the maximum postdoc term. So he fired all postdocs who had
been there for more than three years. That included me, so I had to go out
and find a job.

Back then, job availability in high-energy physics was cyclic. There would
be a lot of them and then there wouldn't be many. This was a time when
there weren't many. I did have a few good opportunities, but they involved
moving away from the Bay area and I didn't want to do that. So Frank
Solmitz, the physics--statistics guy, came up to me one day in the hallway
and said, ``There's a position at SLAC leading a computer science research
group and they were asking me who might be a good computing physicist for
that and I mentioned your name. Are you interested in exploring it?'' I
thought it wasn't really for me but I could explore it. So I went down and
I interviewed. First I interviewed with all the directors and all the group
leaders at SLAC, then I interviewed with all of the professors in the
Computer Science Department on campus. Originally they wanted to get a
famous computer scientist to run that group, but they couldn't find one
that they liked and who liked them, so they decided to get a computing
physicist, which is why they landed on me.

After I returned from interviewing I figured that was it. It was a fun
experience, but I didn't think I wanted it and they didn't want me. Then I
got a call a week or so later saying, ``There's been more than a little
interest in you. What do you want to do?'' I said, ``I think I'd better
talk to the people in the group before I do anything else.'' I went and
talked to the people in the group. They were really good people, so I
thought, \textit{Why not}? So I went down to SLAC to lead this computation
research group. It was set up by Bill Miller, who initially established the
computing facility at SLAC. They wanted him to build up the Computing
Center so he would only come under certain conditions. One condition was
that he be made a professor in the Computer Science Department. Another
condition was that he would be able to have his own computer science
research group at SLAC. SLAC had a lot of physics research groups but he
would have his in computer science and that was this group. He eventually
became Provost of the University (Stanford University), so that position
was open and that's where I went.

\textbf{NF:} How were things set up?

\textbf{JF:} He had a lot of bright people there. A number were in computer
graphics, which was in its infancy in those days. He had set up a really
state-of-the-art computer graphics facility, including movie-making
equipment worth millions of dollars, which was a lot of money in those
days. It was really state of the art. There were people doing research in
other areas of computer science, and a few pure service types doing
job-shop programming for the physicists at SLAC; overall, about ten people
in the group.

\textbf{NF:} So you had the sort of technology advantage that the Bell
Labs' statistics group had rather later on with their workstations.

\textbf{JF:} Yes, this was a fantastic facility. Also, SLAC was a physics
lab and high-energy physics labs had more computing than anybody else
except for weapons laboratories. I had access to the computing facilities
at SLAC, including their mainframe computing system. Very few statisticians
had access to that kind of computing at that time or even fifteen years
later.

\textbf{NF:} What did the job involve?

\textbf{JF:} The job involved mainly running the group as an administrator
and then doing my own research. I~think they expected me to do half and
half: I did about one quarter administration, three quarters research. I
arrived there in early 1972, commuting from Berkeley for the first six
months. Also, I was asked to teach an elementary computer literacy course
in the Computer Science Department. It was a course on algorithms, data
structures and computer architecture. I knew some of those things a little
bit, but in order to teach the course I had to learn them all in detail. It
was one of the most valuable courses I've ever taught in terms of what I
learnt. I still use most of it in my work today.

The research that I wanted to do was in pattern recognition. Even when I
was a student and then a postdoc at Berkeley, I was interested in data. I`d
written some analysis packages, I'd done Monte Carlo, and I'd written a
program to do maximum likelihood. My interest in data worked out well
because most other physicists were more interested in building new
equipment at that time, whereas I was interested in analyzing the data and
that is what got me into computers. I loved computers.

\textbf{NF:} What did you try to do with pattern recognition?

\textbf{JF:} It was called pattern recognition then; it's called machine
learning now. Sort of basic pattern recognition, like nearest-neighbor
techniques. I'd read the Cover and Hart (\citeyear{6}) paper and I was interested in
clustering and in general statistical learning, but it wasn't called that
then. The closest name then was ``pattern recognition.''

\textbf{NF:} Finding groups in data?

\textbf{JF:} Yes, finding groups in data, using data to make predictions,
that kind of thing. I didn't have a clear-cut research agenda at that
particular time. I was just leaving Berkeley where I'd mainly done physics
except for the other sort of statistical things, so I hadn't really
developed a research agenda. I'm not sure I ever had one.

%f2 #&#
\begin{figure*}

\includegraphics{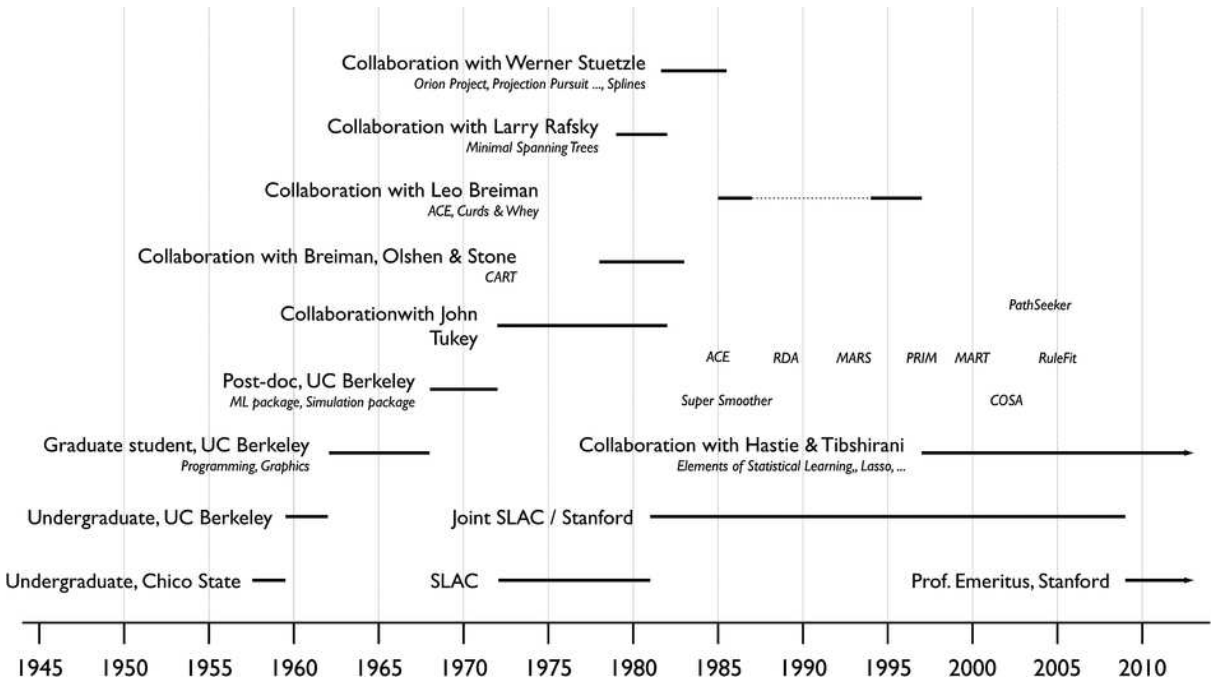}

\caption{An approximate time-line for some of Jerry's major areas of
research and research collaboration.}%\label{fig2}
\end{figure*}

\textbf{NF:} I understand that the group that you were involved
with there had some extraordinary people.

\textbf{JF:} Yes, it did. When I came, it was common then, and may still
be, that in the (Stanford) Computer Science Department professors were paid
half their salary from the Department and expected to go out and raise the
other half externally. One way they could do that would be to work in other
places. In our group we often had computer science professors working part
time. When I came, Gene Golub was halftime in the group. And we had two
visionaries, Harry Sahl and Forrest Baskett. Harry was there when I came.
Forrest joined later. This led to some remarkable developments. (See the
anecdote \textit{Building the first Graphics Workstation} in
\cite*{9}.)

\subsection*{Collaborating with John Tukey 1972--1980}

\textbf{NF:} Just after you moved to SLAC you started collaborating
with John Tukey.

\textbf{JF:} Yes, my predecessor, Bill Miller, was close friends with John
Tukey, so he'd invited Tukey to come out during his sabbatical because, as
we all know, John was very interested in graphics and he was especially
interested in motion graphics. Our facility was one of the very few places
you could do motion graphics. When I arrived at SLAC everyone was excited
that this guy was coming, not because he was a great statistician, but he
because he was well known in computer science for having invented the Fast
Fourier Transform. They were really excited, and I'd never heard of him.

\textbf{NF:} So when John came up you did not actually have a
research project in mind?

\textbf{JF:} No. I talked to him and he told me what he was doing, what he
was interested in, and I found it very interesting. We just hit it off. He
worked on the graphics, I worked a little bit on the graphics but not a
lot. I would watch what they were doing with the graphics---rotating point
clouds and isolating subsets, saying, ``Okay, let's just look at these,''
and so on---trying to visually find patterns in data. John was mainly
working with a programmer in our group.

\textbf{NF:} John never programmed, himself?

\textbf{JF:} Not to my knowledge, at least not code that ever ran on a
computer. He wrote out his thoughts in a kind of pseudo-Fortran, but he
never actually sat in front of a terminal to execute code, as far as I
knew. (See the sample of Tukey's research notes in \cite*{9}.)

\textbf{NF:} What sort of ideas was he having at that time, point
cloud rotation and so on?

\textbf{JF:} Well, if you see the PRIM-9 movie, that's the product and
those were the ideas he had. It was basically integrating the idea of
rotating point clouds in arbitrary orientations. He was very interested in
human interfaces and he developed some really slick controls, especially
given the crudeness of the equipment he had to work with. I was watching
what he was doing and he would iterate to an interesting picture and so I
started to think: \textit{What makes the picture interesting?} and I would
discuss this with him. He said, ``It seems that the pictures we like the
most are the ones that have content; they have a lot of small inter-point
distances but then they expand over the whole thing.'' When I was at
Berkeley I had been working on optimization algorithms and I thought,
\textit{well, what if we defined some index of clumping and then tried to
maximize it with an optimization algorithm?} That was basically the
beginning of projection pursuit and we interacted on that. So I was off
doing the analytical algorithm and John was doing the graphics.

\textbf{NF:} What was John's interest here? He wasn't actually
trying to tackle a scientific problem to do with physics?

\textbf{JF:} Well he thought it would have a big application in physics
because physics has inherently high-dimensional data with a great deal of
structure. It wasn't like the sort of diffuse data that comes from the
social sciences: data from physics have a very sharp structure. In fact, I
think the data set that's illustrated in the movie is a high-energy physics
data set. So his vision was that it could be used for high-energy physics,
but I think he was certainly thinking about the bigger picture.

%f3 #&#
\begin{figure*}
\begin{tabular}{cc}

\includegraphics{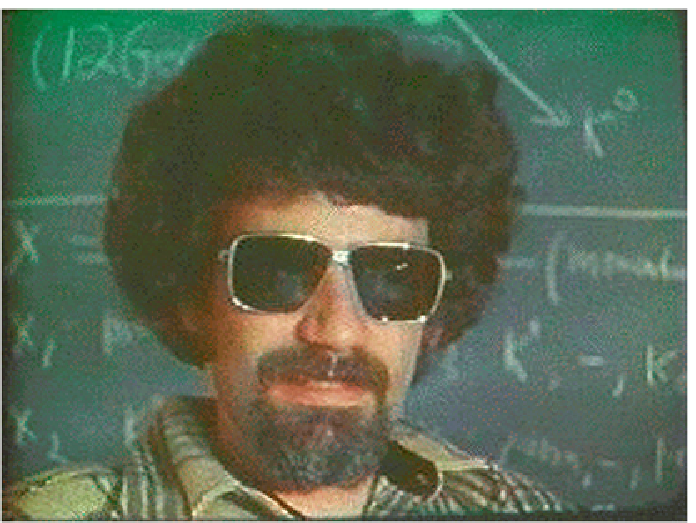}
&\includegraphics{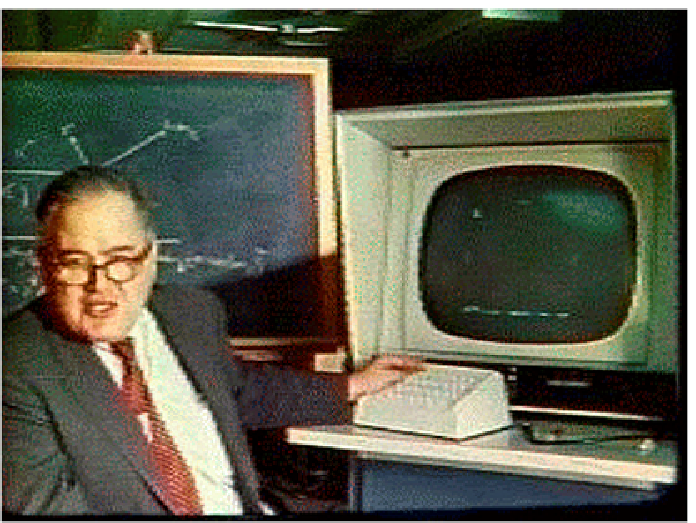}\\
(a)&(b)
\end{tabular}
\caption{Frames from the PRIM-9 video. (\textup{a}) Jerry Friedman. (\textup{b}) John
Tukey sitting in front of the PRIM-9 hardware and using the blackboard to
give his explanation of the variables in the particle physics data.}%\label{fig4}
\end{figure*}

I think he was there four months. When he came back later for a little
while, I said, ``John I think we ought to make a movie of this,'' since we
had a lot of movie-making equipment. My predecessor Bill Miller was a
genius at raising money. He had a graduate student who was interested in
graphics. The student was very smart and wanted the best of everything, so
he got the best of everything. He knew how to handle the movie equipment,
so he made the film just pointing a camera at the screen with John there
talking. So then we had a film\ldots and then no one wanted to edit it.
A~new member, Sam Steppel, had just joined the Group and I asked, ``Would you
like to do the editing?'' And he said, ``Oh yeah.'' It turned out to be a
big job. Anyway, that was the result of John Tukey's first trip to stay
with us at SLAC. We stayed in contact throughout the 1970s and he came back
again for his next sabbatical seven years later.

\textbf{NF:} How did you find interacting with him on the original
Projection Pursuit paper?

\textbf{JF:} He was very full of ideas and he was very stimulating. We
seemed to talk the same language, to think about things the same way. His
approach was operational: here's the task, here's the problem, how do we
approach it, how do we get it done. He didn't seem to be interested in
fundamental principles; he probably was, but he never said so.

\textbf{NF:} A very engineering approach.

\textbf{JF:} Very engineering, that was always his approach. He always
delighted in slightly puzzling you by hiding, not telling you the
fundamental reason for whatever he was doing, what lay behind it, what were
his reasons. He would come to you and say, ``Okay, here's a procedure: you
do this, then you do this, then you do this, then you do that.'' I was
young and brash at the time so I would say, ``John, okay, I understand
that, but why would you do this and this? Why is that a good idea?'' He
would repeat, ``Well, you do this, then you do this, then you do this, then
you do this,'' and I'd say, ``John, but why?'' It would go back and forth
like that, him acting like there was no guiding principle. I guess I was
persistent enough that he would finally get exasperated and say, ``Oh
well,'' and lucidly enunciate the guiding principle; he had it all the
time, he just didn't want to reveal it, at least not right away. His main
thought was he would evaluate a procedure by its performance, not by its
motivations. He wasn't interested in: \textit{Is this a Bayesian procedure
with a particular prior? Is this a procedure that's optimal in some sense?}
He didn't come from that perspective. He would say, ``All right, you've got
a procedure, tell me the operations on the data, the explicit operations. I
don't care where it comes from, I don't care what your motivation is; you
tell me the operations that apply to the data and I'll tell you whether I
think it's a good idea or not.'' That's the way he thought about things.

\textbf{NF:} Do you think he was mentally checking this against a
hidden set of principles or seeing how it sat with his instincts?

\textbf{JF:} I don't know whether he always had a guiding principle or he'd
make one up so I'd stop asking.

I wrote up the first draft of the Projection Pursuit paper, he edited it
and then we discussed it. My first journal publication of any sort in
statistics was the Projection Pursuit paper with John (Friedman and Tukey, \citeyear{29}). This is the only
paper I have ever submitted and had accepted immediately without revision,
and I thought, \textit{This is really neat, I like this field}. But it's
never happened since.

\textbf{NF:} You did some follow-up work with him at SLAC.

\textbf{JF:} Yes, he came back his next sabbatical in the early 1980s. I
think at that time he was on his way to Hawaii because a cousin or somebody
was getting married, and Elizabeth finally convinced him to take a vacation
there on the beach. So he stopped by Stanford and we worked together.

He was very impressed with the fact that the home he was staying in on
campus had a swimming pool; it was the house of a professor who was on
sabbatical. So he wouldn't come in the morning; he would spend his
mornings sitting by the pool, maybe swimming as well, writing out ideas---lots of ideas---about how to analyze high-dimensional data, usually
writing in cryptic words or pseudo-Fortran. Then he would bring them in
later in the afternoon and ask our secretary to type them up. This happened
every day. Later, Werner (Stuetzle) and I would take a look at them and
sometimes discuss them with him.

He finally took off for his vacation in Hawaii and the notes stopped. A few
days later, packages of notes started arriving in the mail, every day
another package from Hawaii. He was thinking on the beach instead of at the
swimming pool.

I've still have many of these notes. After John died, there was an issue of
the \textit{Annals} that had a long article about him by David Brillinger
(\cite{5}), and Werner Stuetzle and I wrote a shorter article
(\cite{27}) talking about his graphics work and our
experiences with him in his graphics work. At that time we thought maybe we
should get the notes together, take a look at them. There are probably a
tremendous number of ideas there that are still revolutionary by today's
standards in terms of data analysis, but this is one of those things you do
when you have time.

\textbf{NF:} Going back to your own personal research, it seems
that it was becoming more statistical.

\textbf{JF:} It was. I was interested in pattern recognition in the general
sense and among the more popular methods of the time were nearest-neighbor
methods and kernel methods. Cover and Hart had shown that, asymptotically,
the nearest-neighbor classification method reaches half the Bayes risk just
with the nearest neighbor. Of course, at that time we didn't appreciate the
difficulty of becoming asymptotic in high-dimensional settings. At the time
people were very excited about it and I thought, \textit{well, if we are
going to use this approach in applications with bigger data sets like those
in high-energy physics, we'll need a fast algorithm to find nearest
neighbors in data sets}. At the time SLAC experiments generated tens of
thousands of observations, not millions like now, but tens of thousands.
The straightforward way to compute near neighbors is typically an
$n^{2}$-squared operation: for each point you have to make a pass over all
the other points. So I started working on fast algorithms for finding near
neighbors, without too much success.

Then I met Jon Bentley, a student of Don Knuth's. He had some really clever
ideas based on what he called \textit{k-d} trees, and so he and I started
working together with another student, Raphael Finkel, on trying to develop
fast algorithms for finding near neighbors. So probably one of the papers
that I am best known for outside statistics is that paper: fast algorithms
for finding near neighbors (\cite{20}). Then Jon
went off to graduate school at the University of North Carolina. After that
he went on to do great things and became very famous in computer science.
The whole \textit{k-d} tree idea is considered a very important development
in computational geometry, and John invented it, an unbelievably bright
man.

Another interesting aspect is that that's what got me into decision trees,
because the \textit{k-d} tree algorithm for finding nearest neighbors
involved recursively partitioning the data space into boxes. If you wanted
to find the nearest neighbors to a point, you'd traverse the tree down to
the box containing the point, find its nearest neighbor in the box and
backtrack up and find its nearest neighbors in other neighboring boxes
using the tree structure. That was the algorithm. I was thinking: okay, if
you want to find nearest neighbors, that's fine, but suppose the purpose of
finding the nearest neighbors is to do classification, maybe there would be
modifications to the tree-building that would be more appropriate for
nearest neighbors in that context. So it occurred to me that in the
nearest-neighbor algorithm you could recursively find the variable with the
largest spread and split it at the median to make boxes. Why don't we find
the variable that has the most discriminative power and split it at the
best discriminating point? So I came up with that paradigm to find the
nearest neighbors. Then it occurred to me that you didn't need the nearest
neighbors at all; you could use the boxes (terminal nodes) themselves to
perform the classification.

\textbf{NF:} When was this happening?

\textbf{JF:} Probably around 1974, before I went to CERN.

That was my initial thinking about what eventually became CART: it came
from the recursive partitioning nearest neighbor algorithm to get the tree
structure. Somewhat later I joined with Leo Breiman, Richard Olshen and
Chuck Stone who had been independently pursuing very similar ideas.

Oh, I forgot to mention that when I first joined the Computation Research
Group in the early 1970s, Gene Golub came to me one day and said, ``I'm
going on sabbatical next year, which means that I won't be here and I'm
worried that if you have an empty position for a year it might not be there
when I get back. So I think you should fill it with someone and I know just
the ideal guy. His name's Richard Olshen and he's in the Statistics
Department.'' So I hired Richard half time. That was in the early days
that I was working on trees. I~was talking to Richard and he asked, ``What
are you doing?'' ``Well, I'm working on this recursive partitioning idea.''
Richard got very interested in it and he has made great contributions to
tree-based methodology over the succeeding years.

\subsection*{Visit to CERN 1975--1976}

\textbf{NF:} After a few years at SLAC, you decided to take a
sabbatical at CERN. Did you have a family at this stage?

\textbf{JF:} Yes, I had a wife and a three-year-old daughter at that time,
and we all went to CERN, in Geneva. It was natural that when physicists
took a year off they went to CERN. It wasn't an official sabbatical, I just
decided I wanted a year away and so I asked for a leave of absence. I was a
staff member, but I wasn't a faculty member. Intellectually, it was not
super stimulating. I was in the computer group which was called Data
Handling and it was a big group at CERN that had the computers. The
professional thing I did was to work on adaptive Monte Carlo algorithms.
What I mainly did was eat their food, drink red wine and dine at a lot of
Michelin three-star restaurants, which is what I mainly remember. CERN was
a lot of fun. SLAC was quite an intense place, whereas CERN was much more
laid back at that time.

\textbf{NF:} Did you visit any other groups while you were at
CERN?

\textbf{JF:} Yes, I did, which turned out to be very important for me. When
I was at CERN I got a letter from John Tukey saying, ``There's this fellow
I know in Zurich at ETH, Peter Huber; he is interested in these projection
pursuit kinds of stuff. You should go and visit him.'' So I went to Zurich
and found my way from the train station to ETH. I'd never met Peter or
anyone else from ETH, so I was standing there in a hallway, and a guy came
up to me and asked, ``Can I help you?'' I guess he knew I spoke English,
maybe it was written all over me. I said, ``Yes, I'm trying to find Peter
Huber.'' He turned out to be Andreas Buja, who was Peter's student at the
time. On that trip I also met another of Peter's students, Werner Stuetzle.
We had a strong collaboration throughout the early 1980s when he came to
SLAC and Stanford. I think Andreas also visited SLAC a couple of times.
Both are unbelievably smart guys.

\subsection*{Interface Meetings}

\textbf{NF:} Returning to your time at SLAC, you'd started
attending Interface conferences and meeting people\ldots

\textbf{JF:} Yes. I met Leo Breiman and Chuck Stone at an Interface meeting
in 1975. Leo gave a talk about nearest neighbor classification or something
and I was working on these fast algorithms at the time, so I raised my hand
at the back of the room and said, ``We've been working on some new fast
algorithms for finding nearest neighbors.'' After the talk Leo looked me
up. He was very interested and we started talking, but that was pretty much
it. But then he sent me a letter while I was at CERN saying that he was
organizing a meeting in Dallas in 1977; he called it a conference on
\textit{The Analysis of Large and Complex Data Sets}. Leo was another
visionary; he saw into the future of data mining. He invited me to give a
talk there. I'd never been to Dallas and so soon after I got back I went to
that meeting, and that meeting to a large extent changed my life
professionally. I met Larry Rafsky there, with whom I later collaborated,
and I also met Bill Cleveland.

\textbf{NF:} How did this conference change your life?

\textbf{JF:} Because I met Leo again.

\textbf{NF:} We'll talk about Leo shortly. You did some work with
Larry Rafsky around this time.

\textbf{JF:} Yes, we started talking about some of our mutual interests in
computational geometry (near neighbors). This led to the work in the late
1970s, early 1980s on using Minimal Spanning trees for multivariate
goodness of fit and two-sample testing, leading also to general measures of
multivariate association. Two \textit{Annals} papers came out of that
(Friedman and Rafsky, \citeyear{24}, \citeyear{25}). I was also refining the recursive
partitioning idea, extending it in various ways, and I worked with Larry a
bit on this as well. He was a very bright guy with lots of ideas. I learned
a lot from him.

\subsection*{CART and Leo Breiman 1974--1997}

\textbf{NF:} Let's bring the background murmurs about recursive
partitioning to the foreground and talk about CART. How did this celebrated
collaboration come about?

\textbf{JF:} After Larry and I wrote the two papers using minimal spanning trees, we
started working on the CART idea. Richard Olshen was at UC San Diego at
this time (mid-1970s) and he made trips every once in a while back to
Stanford, and he would come out and visit me at SLAC. Sometimes I would
tell him about the more recent work on trees. He'd done some nice
theoretical work with Lou Gordon (Louis I. Gordon), a former Stanford
professor who was working in industry at that time. I told him how we were
extending decision trees and he said, ``It sounds a lot like what Leo
Breiman and Chuck Stone are doing down in LA.'' He tried to explain to me
what they were doing and I didn't quite get it; and apparently he was
trying to explain to them what we doing and they didn't quite understand
either. Finally, Chuck called me and we had a long discussion. We'd been
working totally independently, but there was a huge amount of commonality
in what we were doing. So I guess it was Leo who finally suggested that we
have a meeting down in southern California. They were both consultants for
a company called Technology Service Corporation that was operating on
government contracts, mostly environmental things I think. Leo was
basically a full-time consultant there and Chuck was also a consultant. In
fact, some of the technical reports that they wrote then are the classic
articles on trees. So, Larry and I and Chuck and Leo, we went down there
(Richard wasn't there) and had a meeting at TSC. We talked about how very
exciting it was and that there was a lot of commonality in our respective
approaches. There were some differences, and we discussed which ones seemed
best. Then Leo said, ``Hey, I think we ought to write a monograph.'' We
would never get something like this published in a statistics journal (of
the day). So we set off to write it, and that's how the monograph was born
(\cite{4}).

\textbf{NF:} As I recall, there was other work on recursive
partitioning going on about this time.

\textbf{JF:} Well, it's one of those ideas that's continually re-invented.
Everybody who re-invents it thinks this is their ``Nobel Prize'' moment.
There was the work of \citet{32}, in the early 1960s at the
University of Michigan Social Science Center; they did trees. Then there
was Ross \citet{35} who was doing what he called the Iterative
Dichotomiser 3 (ID3) algorithm, a crude tree program, at about the same
time. Later he did C4.5, which turned out to be very similar to CART,
although there are a few differences. We take pride in the fact that CART
came ten years earlier than C4.5, but it was Quinlan and the machine
learners who popularized trees. We did CART and it just sat there:
statisticians said, ``What's this for? What do you do with it?''

\textbf{NF:} And you'd also implemented the software and made it
available.

\textbf{JF:} Yes, we'd made it available. Then we got the idea of trying to
sell it and that's how our little company got started.

\textbf{NF:} First, let's talk about your long collaboration with
Leo. This was the beginning.

\textbf{JF:} Right, it started with CART because we were trying to write
the software. I had written the initial software, but Leo had a lot of good
ideas about what should be in it and how it should be structured, the user
interface etc., and so we were collaborating on that. In the
meantime, Leo left UCLA and became a full-time consultant.

\textbf{NF:} He was a probabilist at one stage.

\textbf{JF:} He was a probabilist, he used to say \textit{probobobilist}.
Then he came back to academia in 1980 and joined the Statistics Department
at Berkeley and, at the same time, Chuck came up to Berkeley.

\textbf{NF:} Would you say Leo was an unusual appointment at
Berkeley for that time?

\textbf{JF:} Yes. He had solid mathematical credentials. He was like Tukey
in this sense: he could do this super empirical stuff but he was also very
strong in math, so they couldn't say that he was doing methodology because
he couldn't do math. I have no idea why they hired him, but my guess would
be they wanted to start getting into the computer age and they brought him
in. He bought their first computer, a VAX, installed it, and did its care
and feeding for a long time, so it was an incredibly wise appointment from
that perspective, as well as many others.

\textbf{NF:} How did the collaboration go?

\textbf{JF:} We'd started the collaboration with CART, we'd decided to
write the book, and we'd parceled it up into different parts. Then Leo
says, ``If we write this program called CART and decide to sell it and we
sell a thousand copies at a hundred dollars each, you know how much money
that is?'' So we decided, okay, we would form a company, California
Statistical Software, and try to sell CART. So we had to have a product.
Leo was at Berkeley at that time, so we started a pattern that persisted
for roughly the next ten years. Every Thursday I would go up to Berkeley. I
would leave here around 10 am, get up there around 11 and park on Hearst
Avenue. Leo would block out the whole day; nobody else would come to see
him for that day. We would go to his office and start working. Around noon
he'd say, ``Jerry let's go have lunch,'' So we'd go over to the same place
every time, a cr\^{e}pe place over on Hearst Avenue. We'd have usually the
same spinach cr\^{e}pe with sour cream, and an espresso. Then we'd go back
to his office and work, punctuated with me running out to feed the parking
meter on Hearst. It was all conversational; we weren't sitting there
writing or typing into a computer, we were just discussing the whole time.
Typically, around 5.30 or when the progress seemed to be slowing, Leo would
say, ``Jerry let's go have a beer,'' so we'd go down to \textit{Spats},
which is a pub on Shattuck Avenue. After we'd had a few beers Leo would
say, ``Jerry let's go to dinner,'' so we'd go to one of Berkeley's better
restaurants and have a nice meal. Then Leo would go home and I'd drive back
down to Palo Alto. That was the routine every Thursday for a very long
time.

\textbf{NF:} What was his approach to problems?

\textbf{JF:} He was like Tukey: ``Don't tell me the motivation, tell me
what you do to the data.'' He was totally algorithmic. There was no obvious
sort of fundamental principle like: \textit{This is a Bayesian procedure
with a particular prior}. It was never that kind of thinking, starting from
any kind of guiding principle; it was just what it made sense to do with
the data.

\textbf{NF:} Would you categorize this as the computer science way
of tackling data rather than the statistical way\ldots?

\textbf{JF:} I would have then.

\textbf{NF:} \ldots in the sense that what you are doing is looking at
a specific data set and you don't know whether what you've done is going to
work on any other data set?

\textbf{JF:} Well, we generally weren't working on specific data sets, we
were trying to develop methodology for classes of problems. It was like
developing CART: CART could be used on a wide variety of data sets, so
could ACE (Alternating Conditional Expectation) (\cite{2}), so could Curds and Whey
(Breiman and Friedman, \citeyear{3}). We were
thinking methodologically. In other words: \textit{Problem. I've got data,
there's an outcome, there are predictor variables, the data is of a certain
kind. Now how do we make a procedure that can handle this problem?} I don't
think we ever actually analyzed a specific data set together, except for
examples that we used in papers to illustrate the methodology. Analyzing a
data set where the interest was not in how well the method did but in the
answer that you got from the data set, we both did a lot of that as well.

\textbf{NF:} What motivated ACE?

\textbf{JF:} The idea was to simultaneously find optimal transforms. There
were all these heuristics and rules for transforming data in the linear
regression problem: do you take logs, or do you take other kinds of
transformations? In fact, I think Box--Cox was a sort of automated method
for trying to find transformations from a parametric family of functions.
We were involved with smoothers, so we thought about how we could
automatically find good transformations without having to restrict them to
be from a parametric class of functions, just see if you could estimate an
optimal set of transformations.

\textbf{NF:} ``Optimal'' in what sense?

\textbf{JF:} Optimal in the squared error sense\ldots  of course under a
smoothness constraint, otherwise there were an infinite number of
transformations that would fit the data perfectly. So you had to put in a
smoothness constraint, which we did explicitly by using smoothers in the
heart of the algorithm. I remember one of the Thursdays when I went up to
Berkeley, Leo asked me, ``If I have two variables, how do I find the
function of one of them that's maximally correlated with the other one?'' I
said, ``Well, if you do a smooth, you take the conditional expectation of
one of them given the other one, okay?'' That doesn't necessarily maximize
the correlation, so we started thinking: \textit{Okay, what if we did it
one way and then, given that curve, smooth that against the other one?}
Later we went back to Leo's house where he had an Apple 2. He programmed it
in Basic, just the simple bivariate algorithm. He simulated data from a
model where the optimal transformation in both cases was the square root.
The Apple 2 was not a very fast machine, so we could watch it iterate in
real time, displaying the current transformations at each step. Starting
from linear straight lines, we saw the transformations begin to become more
and more curved with each iteration until they converged. It was an
exciting moment for us.

So we developed that idea and then Leo got very excited about the theory.
He never took theory very seriously but he loved to do it, so he looked at
the asymptotic consistencies and things like that, and we had a great
time.

In the early 1990s I went on sabbatical for a year and we didn't
collaborate then, but it picked up again in the mid-1990s. Leo called me
one day and just said, ``Jerry, I'd like to work with you again,'' and we
didn't even have a specific project to work on. I went up to Berkeley and
we kicked around what we could work on. I said, ``Well, one problem I've
been churning in my head but haven't gotten very far on is multivariate
regression, where you have multiple responses.'' So we started kicking that
around and that led to the Curds and Whey paper, which was a Discussion
paper at the Royal Statistical Society.

This collaboration wasn't quite in the same mode as before. I wouldn't go
up to Berkeley nearly as much because infrastructure had developed so that
it was possible to work apart productively and so we basically did it
through e-mail. The idea was motivated by my familiarity with PLS (Partial
Least Squares). PLS had a mode where it had multiple outcome variables as
well as multiple predictor variables. The one-outcome-variable case was
just a special case. In the work that I had done with Ildiko [Frank] to try
and understand PLS (see below), we only treated the single outcome case.
I~wanted to try to understand the multiple outcome procedure to see if one
could find a more statistically justifiable approach. So Leo and I worked
on that together and that was great fun.

In this paper we reversed roles. Generally, in our collaborations I
concentrated on the methodological part and the computing. Leo would
usually do the theory. In this paper our roles were reversed: Leo wrote the
program, Leo had the data, and I worked out the theory.

\textbf{NF:} Why Curds and Whey?

\textbf{JF:} I'll retell the story I told in the memorial article I wrote
about Leo. I came up with the name ACE. I liked it a lot but Leo hated it,
absolutely hated it. This was one of the afternoons after we finished and
we'd gone down to \textit{Splats} for a beer and we were still discussing
this. Leo didn't like it and I liked it, so we were going back and forth.
And then out of nowhere Leo said, ``Okay Jerry you've got it, it's ACE.''
It was most unusual for Leo to yield so easily. He usually stuck to his
guns and so did I. I looked at him in a puzzled way, like, \textit{That was
too easy}, and he said, ``Look across the street,'' so I looked across the
street and there was a hardware store with this big red sign, \textit{Ace}.
When we gave the invited \textit{JASA} paper in 1987, Leo brought a bunch of bags
from Ace Hardware that had a big ace on them and distributed them around to
the audience. Later on when we did the multiple response multivariate
regression work, we had another argument about how to name that procedure.
Leo proposed Curds and Whey, which I really didn't like, but I felt that
since he had conceded on ACE I would concede on that. It was Leo's thinking
about the fact that we were separating a signal from the noise, the good
stuff from the bad stuff, separating the curds from the whey or the other
way around, I guess, in cheese manufacturing.

That collaboration was a couple of years, maybe three years. I think to
some extent our interests separated at that time. They tended to be
concerned with very similar problems. He did the nonnegative garrotte and
then got into bagging and I was getting into boosting at that time working
with Rob [Tibshirani] and Trevor [Hastie]. Both approaches were based on
ensembles of trees, but from different perspectives. I~knew what he was
doing, but we didn't have constant interaction and involvement. When we got
together we always had a good time.

\textbf{NF:} Talking about Leo has had us leaping through the
decades. Let's return to the period when you were still full time at SLAC.
Had you met anybody from the Statistics Department at this stage?

\textbf{JF:} No, not at this stage. I didn't start interacting with the
Statistics Department until the late 1970s.

%s4 #&#
\section{The move to Stanford University}

\textbf{JF:} I was hanging out around the department for seminars, but I
had no official position. So Brad [Efron] asked me to teach a course.

\textbf{NF:} Did you think you were doing statistics?

\textbf{JF:} Well yes, I knew the stuff with Rafsky was statistics, it was
hypothesis testing. That's what I taught in the course. It's probably as
close as I've come to classical statistics. The minimal spanning tree was
not classical statistics but the rest of it was.

That brought me closer to the department. While I was at SLAC I wasn't on
the faculty there. I was just a staff member, which meant I couldn't write
proposals and submit them to NSF or other agencies, Department of Energy,
or others who might sponsor my kind of work. SLAC was sponsoring it and
that was wonderful, but sometimes I really could have used a little more
money to do things. So I wanted to write proposals and for that I needed to
be some kind of professor. Paul Switzer was Chair of the department at that
time, so I went to him and said, ``Is there any way you could make me
something like a consulting professor of the Department, some official
thing? This will allow me to write grants and reports on behalf of Stanford
University.'' He said, ``Okay, we'll try it.'' So all the paper work was
gotten together and submitted to the administration, letters and
everything. It came back and Paul said, ``Sorry we can't do it. We're not
making any more consulting professors; there is some political thing going
on that has nothing to do with your case, but they are not doing consulting
professors. However, they did say that your folder looked pretty strong, so
why not try for a regular professor?'' And so Paul did and it worked. Paul
probably did the lion's share of the work on it because he was Chair.
That's how I became a professor.

\textbf{NF:} As well as having a job at SLAC?

\textbf{JF:} I became a half-time professor and half-time at SLAC instead
of full time.

\textbf{NF:} So this was effectively your formal entry into the
statistics community. Did you find yourself welcomed? Here's mainstream
statistics flowing along and this guy surfs in on a wave from a merging
stream with no statistics background whatsoever, but with lots of skills
and different ideas about how to approach data. Was this a great issue for
you?

\textbf{JF:} Yes in general, but certainly not at Stanford because they hired me. I
always felt very welcome in the Department. But I don't think the more
general statistics community understood what motivated me. I recall once
Colin Mallows listening to one of my talks, and he said afterwards, ``Boy
this is really fascinating, but it's not statistics,'' and I think that was
the general feeling, that what I was doing was perhaps interesting but not
statistics. Where's the math? Where are the usual trappings of research in
statistics? It really wasn't that sort of stuff, with the possible
exception of the minimal spanning tree work. So in that sense, I don't
think there ever was any hostility of any kind, just that people were
puzzled: how was what I was doing related to statistics?

\textbf{NF:} And yet what you were really doing was what you
described earlier: you and John Tukey thinking the same way, you'd have an
idea about how to attack something and you'd see how it worked on the data.
Your work wasn't being informed by fundamental principles\ldots  or
was it?

\textbf{JF:} I think that had more of an influence on me many years later,
and John thought I'd sold out. He really thought I was trying to think
about fundamental principles, whereas I was developing things and using
elegance of the algorithm as a criterion. John had a real distaste for
that.

\textbf{NF:} Do you feel that you had developed some sort of a
canonical way of tackling the sorts of problems that you approached?

\textbf{JF:} Probably, but I can't think of it right now. I operate in the
model of a problem solver: here's a problem, I have a certain set of tools
and skills that I use, and so that directs everything. Probably there is a
great deal of commonality simply because my skill set is limited, but I
don't think I consciously think that way.

\textbf{NF:} Suppose a young person came to work with you and you
treated that person the same way as John Tukey used to treat you: you do
this, you do this. If you got pushed would you make up a principle or would
you actually be able to find a principle? You suggested earlier that maybe
John made the principle up to shut you up.

\textbf{JF:} A heuristic principle perhaps, I don't think I could come up
with a deep theoretical principle, or maybe I could if I thought about
it.

\textbf{NF:} Joining the department put you into contact with
mainstream statistics and statisticians and you started going to more stats
conferences? How was being in that department changing what was
happening?

\textbf{JF:} Well, I started becoming more conscious of statistical
principles. I don't think it changed the way I approached problems a lot. I
recall a statement of John Rice's when he was asked whether he was a
Bayesian or a frequentist and he said, ``I'm an opportunist.'' And that's
how I view it: \textit{Here's a problem. How do we solve it?} I will try to
attack the problem from any direction I'm capable of.

\textbf{NF:} You were also coming into contact with a remarkable
group of statisticians in the department, who were doing extraordinary
things.

\textbf{JF:} I think subconsciously that really shaped my thinking a lot.
That's maybe why Tukey thought in later years I was selling out. I did
think about principles; I~think they were in the back of my mind, informal
principles that I didn't apply formally.

\textbf{NF:} Did John ever visit you once you had moved into that
department?

\textbf{JF:} Yes, oh yes, at least a few times. I do remember one time we
were driving along Campus Drive and I said, ``You know, John, now I'm in a
statistics department and officially in statistics, maybe I should really
go and learn basic statistics, theoretical statistics, all the usual
stuff.'' John looked at me and went: (\textit{raspberry sound}). Whenever
you said anything to John, presented an idea or whatever, John didn't tend
to lavish praise, that wasn't his style. So if he sat still and listened to
you quietly, you knew he really liked it. If he had doubts about it, he
wouldn't say anything, but you'd see his head going slowly back and forth;
and if he really didn't like it, he'd interrupt you by giving a thumbs down
and blowing a raspberry. So that's what I got when I asked him whether I
should learn statistics. I'm not sure he was exactly right and over the
course of the years I did learn some traditional statistics with the help
of my friends, colleagues and students, which I think helped me a lot.

\subsection*{The Orion Project}

\textbf{NF:} You developed more strong collaborative work at
Stanford. What was the first one?

\textbf{JF:} Around 1981, the department had an opening for an assistant
professor and I think Werner [Stuetzle] had just got his degree. I said,
``I know this really smart guy that I met at ETH. I think I can pull it off
so we pay him half time with my group, do you want to hire him?'' They
thought about it and, to cut a long story short, they said, ``Sure.'' I
convinced my bosses at SLAC that we could do it, so we hired Werner half
time at Stanford and I had Werner half time in my group at SLAC. I~started
my collaboration with Werner, which was very profitable intellectually and
great fun over the years.

\textbf{NF:} What sort of things were you doing?

\textbf{JF:} Well, we started a graphics project. Werner had worked with
Peter Huber on graphical techniques, as Peter was very interested in that.
So we got some money from the Office of Naval Research and started to put
together a graphics workstation. We felt: it's been ten years since PRIM 9,
the technology has advanced dramatically, let's see what we can do now.

So we jointly worked on that; we called it the Orion Project and it was
great fun. (See the anecdote \textit{The Orion Project---building a second
Graphics Workstation} in \cite*{9}.)

\subsection*{Searching for Pattern}

\textbf{NF:} Your full-time work with SLAC had been a very exciting
period of your life. Now you had moved across to the department of
statistics, how long did the interactions with SLAC continue?

\textbf{JF:} They tapered off a little bit because I was only half time
there and running the group was about a quarter-time exercise, so I had
less time to work on SLAC types of things. But it was still very valuable
to be in that group. I still had access to a lot of resources that I
wouldn't have had otherwise.

\textbf{NF:} How did this change your sources of inspiration for
things to work on?

\textbf{JF:} I was always interested in what Leo called large and complex data sets (now
called ``data mining''): data that was collected not necessarily for the
purpose for which you are using it; it has mixtures of all kinds of
variables; the experiment wasn't designed; it was usually observational
data. I guess it's a kind of data that I first encountered in physics,
moderately high-dimensional, a fair amount of data, the number of
observations usually considerably larger than the number of measured
variables. I was always interested in developing general-purpose algorithms
where one could pour the data in and hope to get something sensible out
without a lot of labor-intensive work on the part of the data analyst.

\textbf{NF:} Jerry's search for pattern?

\textbf{JF:} Yes, I guess a generalized pattern search of data, usually
focused on prediction problems.

\textbf{NF:} Looking forward from your arrival at SLAC, first there
was Projection Pursuit where you were looking for groups in
high-dimensional data\ldots?

\textbf{JF:} I think I was associated with four Projection Pursuit papers.
One was the original Tukey paper (\cite{29}), then there was
a regression paper with Werner Stuetzle (\cite{26}), then
I wrote another follow-up paper (\cite{13}) in the original Tukey
style, and one with Werner on density estimation (\cite{28}).

In the mid-1970s I began work on trees which carried through to CART. Then
I went back to trees later in the 1990s when the various ensemble methods
were coming out. Ensembles of trees seemed especially appropriate for these
kinds of learning machines because trees have a lot of very desirable
properties for data mining. Trees just have one problem: they are not
always very accurate. So the ensembles of trees cured the accuracy problem
while maintaining all of the previous advantages; they are very robust,
they can deal with all kinds of data, missing data, and that's the kind of
thing I was interested in: off-the-shelf learning algorithms. You could
never do as well as a careful statistician or a careful scientist analyzing
the data very painstakingly, but it could give you good first answers, that
was the idea. That's basically what drove me.

Research interests are a random walk. You get an idea and you pursue for a
while. It may be similar to what you were working on before or it may be in
an entirely new direction. You work on it for a while until you get stuck
or find something more interesting. I~tend to have these problems that I
would like to solve and can't solve immediately. I put them in the back of
my mind and then when I'm reading or hearing talks or every once in a while
someone says something that may have nothing to do with what's in the back
of my mind, it will trigger something: \textit{Ah ha! There's an idea that
I can try for this problem}. So I go back and work hard for a while; either
I push it a little bit further or I don't but it's still there. I've got
this residual set of problems that I hope to solve some day; sometimes I do
get them solved.

\textbf{NF:} You'd worked on CART with several people, Projection
Pursuit with John Tukey and Werner Stuetzle, and ACE with Leo. Then
what?

\textbf{JF:} Other stuff with Werner, SuperSmoother (Friedman, \citeyear{12}) and a
paper on splines (\cite{22}). Then MARS
(Multivariate Adaptive Regression Splines) came after ACE. It started in
the late 1980s. I wanted a technique that would have the properties of CART
except that it would make a continuous approximation. One of the Achilles'
heels of trees is that they make a discontinuous, piecewise constant
approximation and that limits their accuracy. Also, I'd read de Boor's
little primer on splines (\cite{7}) which Werner showed to me. I'd
learnt most of what I knew about smoothing from Werner. Smoothing was an
important tool and I believe his thesis work had a lot about smoothing.
After that I knew something about splines, so I pieced together the idea.
You can think of CART as recursively making a spline approximation but with
a zero-order spline which is piecewise constant, so I tried extending that
so I could use a first-order spline which was a continuous approximation,
discontinuous derivatives but a continuous approximation, and then you can
generalize the approach to higher orders (although in the implementation I
didn't).

\textbf{NF:} As I recall, this ended up being a very large
paper.

\textbf{JF:} Yes, the MARS paper was 60 pages of description and then there
was another 80 pages of discussion, so it ended up as a 140-page paper
(\cite{16}).

Apart from MARS, I also developed a technique I called Regularized
Discriminant Analysis (RDA; \cite{14}). Some of my work was inspired
by work that was going on in chemometrics. There was a technique they
called SIMCA, which was basically a strange kind of quadratic discriminant
analysis, viewed from a statistical perspective. It's an acronym for Soft
Independent Modelling of Class Analogies (\cite{40}). That
was used a fair amount for classification problems in chemometrics.

\textbf{NF:} I recall a meeting involving some chemometricians
where you and Ildiko presented a paper on your views about PLS, where you
showed that it had some significant deficiencies. Have your views on this
subject ever been accepted by the chemometrics people?

\textbf{JF:} I don't think so, no. I went to a chemometrics conference two
or three years ago and everything was still PLS after 20 years. In the
machine learning literature everything is a machine, every algorithm is
called a machine. Before that, every algorithm was called a network in
Neural Nets. In chemometrics everything is called some kind of PLS. You
reminded me: Ildiko and I wrote a paper trying to explain PLS from a
statistical perspective (\cite{11}). Also, when boosting came
out much later, Rob and Trevor and I tried to show what it was doing, again
from a statistical perspective. We did PLS and it turns out it's very close
to ridge regression. I don't think the PLS people appreciated it at all.

PLS definitely has limitations. One thing is that if the variables are all
uncorrelated, then it doesn't regularize at all. At least ridge regression,
which is very similar, still regularizes in that kind of situation. So it
depends upon the predictor variables being highly correlated to impose this
regularization, whereas ridge regression, which gives pretty much the same
result for highly correlated variables, also regularizes in the absence of
a high degree of correlation.

\textbf{NF:} And RDA?

\textbf{JF:} RDA related to this SIMCA thing. It was a very simple idea
about linear discriminant analysis and quadratic discriminant analysis. You
consider an algorithm that is a mixture of the two. Then in the second
part, when you do the quadratic discriminant analysis, you regularize the
covariance matrices in a ridge style so there are two regularization
parameters for the two covariance matrices, each being estimated
separately. Each of the separate covariance estimates is blended with the
common covariance, their average, with degree of blending being another
parameter of the procedure. I liked that idea.

We wrote the paper about PLS when I was on my sabbatical in 1992. The
sabbatical was broken up into small pieces, part of which was in Australia.
That's when you and I started working on multivariate geochemical data.

\textbf{NF:} Yes, that led us to PRIM. Would you like to say a
little bit about PRIM (Patient Rule Induction Method)?

\textbf{JF:} The idea there was hot-spot analysis. Data mining was coming
in and one of the things that people wanted to do was look for needles in
haystacks, hot-spots in data, for example, in fraud detection. You expect a
fairly weak signal, but what you hope for is that it's identified by a very
sharp structure in at least a few of the variables. PRIM (\cite{21}) was a recursive partitioning scheme but different from CART
which was very greedy and aggressive. That's where the ``Patient'' comes
in: it was meant to find a good split but only split a little bit and be
patient and then look for another split, that was the idea.

\textbf{NF:} {There was an earlier bias-variance paper in the
1990s}.

\textbf{JF:} Yes. There was kind of a cottage industry in the mid-1990s;
everyone was aware of the bias-variance decomposition of prediction error
for squared error loss regression and it intrigued people to try and
develop something analogous for classification. Here the loss is either
zero or one, and the goal was a corresponding decomposition of the
misclassification risk. There were numerous papers on that. Leo wrote one
(\cite{1}) and there were a lot in the machine learning literature. I
got the impression that you really couldn't find such a decomposition, but
what you could do was look at traditional bias and variance, which are well
defined, and see how those two kinds of estimation errors, like bias and
variance in estimating the probabilities, reflected themselves in
misclassification risk. So I wrote this paper (\cite{17}) where
basically I showed that the curse of dimensionality affects classification
much less severely than it does regression. In regression, things get
exponentially bad as the dimensionality increases, but not necessarily for
many types of classification. So that is why things like nearest neighbor
and kernel methods, which don't work terribly well in regression in
high-dimensional settings, can perform reasonably well with classification:
the curse of dimensionality doesn't hurt them as much. This is especially
so with over-smoothing the density estimate: it can be very severe and can
introduce huge error in the density estimate, but need not introduce much
error in classification.

I didn't know where to publish that paper, or indeed whether to publish it
all. Then a friend of mine, Usama Fayyad, contacted me. He was one of the
early people in data mining and may even have coined the term ``Data
Mining.'' He was starting a journal of data mining. He said, ``I'd like a
paper from you in the first issue,'' so I said okay. I had this one just
sitting there, so I sent it off to him. It turned out---and I didn't know
this until much later---that paper was read by a data mining fellow
in Israel, Saharon Rosset. He felt that this showed that statistics could
contribute to data mining. So he decided he wanted to come to Stanford and
study. He was one of the best students we've ever had. I learned a lot from
Saharon and still do. So I would say the biggest success of that paper was
that we got Saharon to come to our department.

\textbf{NF:} When was it that you had the insight about
high-dimensional data, that every point is an outlier in its own
direction?

\textbf{JF:} That came from Projection Pursuit. Some time in the late
1980s, early 1990s, outlier detection was a big issue for people. I had
seen it in various papers and talks. I thought that it might be a natural
application of projection pursuit. Projection pursuit looks for directions
in the space such that when you project the data it has a particular
''interesting'' structure defined by a criterion that you then try to
optimize. So I thought, \textit{OK, we'll define a criterion that looks for
outliers}. I came up with a criterion, programmed it up, tried it out and
it was working beautifully. It was finding all kinds of outliers and the
nice thing about it is you see the projection. So in that projection here
is the data, here is the point, there is no other inference to be done;
it's an outlier, there it is. I was very excited and after trying it on
data, both simulated and real, I thought, \textit{Well, we've got to
calibrate this. How many outliers does it find when there are none?} I
generated data from a multivariate normal distribution, tried the algorithm
and it found this incredible outlier. I thought, \textit{Okay, that can
happen, it's an accident}, so I removed that point and searched again. It
found another one, another projection with a far outlying point. It just
kept doing this. I~could just peel the data. I~found this was very curious
and I mentioned it to people and I believe it was Iain Johnstone who came
up with the explanation that every point is an outlier in its own
projection. That was the phenomenon that I was just discovering
empirically.

\textbf{NF:} You've mentioned the term ``data mining,'' {which came from a nonstatistical community. What were your interactions
with these other communities?}

\textbf{JF:} In the early 1990s I was becoming aware of the machine
learning field. I was invited to give a talk at a NIPS (Neuro Information
Processing Systems) conference some time in the very early 1990s. That
opened up a different world for me because there were all these people who
were doing things with similar motivations but not with statistics, not in
a statistical mode. They were almost entirely algorithmically driven. I
felt that was wonderful, so I gave a talk there and I went back to those
conferences throughout the 1990s.

\textbf{NF:} Had they been aware of any of your work?

\textbf{JF:} Well, they must have been aware of some of it because they
invited me to give a talk. I don't know how much my work was referenced in
their papers, probably some. It was interesting the progression throughout
the 1990s when I went to those conferences. At the first one I attended
there was lots of discussion of hardware and these were mostly electrical
engineers. In fact, there were two groups: the engineers who used neural
nets and neural-net-type ideas to solve prediction problems; and the
psychologists who used them to try to understand the brain and how adaptive
networks can learn things, the basic learning theory. For the engineering
part it is interesting how it evolved from a concentration on programs and
hardware to looking more and more like statistics. And now it's basically
statistics. They discovered Bayesian methods. I remember in early
discussions with machine learning people I tried to explain why fitting a
training data as closely as possible doesn't necessarily give you the best
future prediction, or what they call generalization error. Now they
understand that completely, but in those days it was a little hard for some
of them to grasp the concept. To be fair, their interest was in very low
noise problems like pattern recognition. Obviously there exists an
algorithm that can tell a chair from a table every time, the brain can do
it, so the Bayes error rate is zero on that. It's just that you can't come
up with an algorithm to achieve the Bayes error rate. Those were the kind
of problems they were interested in. So in that case fitting the training
data as well as possible is the right strategy. If the Bayes error rate is
zero, there's no noise.

\textbf{NF:} There were a number of distinct communities\ldots

\textbf{JF:} Yes. There were three distinct fields, maybe more, that I know
about. There was statistics, there was artificial intelligence and then
there was data base management.

\textbf{NF:} Where did the computer scientists fit in?

\textbf{JF:} Computer scientists were doing data base management and
artificial intelligence. Machine learning evolved, at least as far as I
know, out of AI. Data mining originally emerged out of the data base
management area. It's all kind of a blend now and everyone is learning more
of what the other people are doing. The machine learners and data miners
are learning more statistics and their research is looking more and more
like statistics. Some statisticians are learning more about methodology and
algorithms and their work is looking a lot more like machine learning or
data mining.

\subsection*{Students}

\textbf{NF:} We've talked about one or two students you were
involved with before you joined the Department, but once you joined you had
some formal responsibilities to supervise these students.

\textbf{JF:} Yes, I had a number of students and I enjoyed them all in
different ways. One of the real advantages of being in an academic
department is that you get to be around students with young fresh ideas and
that eagerness that hasn't been stilted by time.

\textbf{NF:} What collaboration did you have with your students?

\textbf{JF:} I certainly collaborated on their thesis work. Probably the
student that I had the biggest and longest collaboration with was Bogdan
Popescu, from Romania.

\textbf{NF:} Your style of doing things clearly influenced a lot of
people who were around you at that time as students.

\textbf{JF:} I think so, yes. Especially in my early days in the Department
my way of thinking about things was really very different; it's not so much
any more. We've got Rob and Trevor and Art [Owen], all of whom were
students when I first came. Art was actually my student. Rob and Trevor
were not officially my students, but they came up to SLAC a lot.

\textbf{NF:} They got infected by what they saw.

\textbf{JF:} Trevor was Werner's student, so he got infected strongly and
so did Rob, I think: the more phenomenological way of thinking, less the
theorem--proof--theorem--proof--theorem--proof approach. Not that I
devalue that approach. I don't want to give that impression, it's just
different. I'm not good at it. I don't have the skill to do it.

%s5 #&#
\section{Stanford---The New Millennium}

\textbf{NF:} So far at Stanford, we've threaded our way through the
1990s and into the first decade of the new millennium and during this
period you have commenced another very significant collaboration with some
of your Stanford colleagues.

\textbf{JF:} That's right. There were some very impressive and interesting
developments in the machine learning field in the late 1990s and also in
statistics. One of them was Leo's bagging idea, which was a very simple but
clever idea. Then there were the boosting ideas that came out of the
machine learning literature that were introduced by Freund and Shapire
(\citeyear{30}). I started to become fascinated by this because it had a similar
flavor to PLS in the sense that it appeared to work reasonably well but it
wasn't clear why. Again, a cottage industry developed as to why. The
machine learners had their own approach using what they called the PAC
Learning Theory (PAC stands for Probably Almost Correct), which was a way
of looking at it which was very satisfying to them. It was a good way to
look at it, but I think we didn't quite understand it. If it's analyzing
data, it's doing what statistical algorithms do, therefore, there should be
some sort of sound statistical basis for it. So Rob, Trevor and I started a
collaboration to try to figure out from a statistical point of view why
this thing was working so well.

It was interesting in the sense that we didn't have the answer when we
started the collaboration. This was similar to working with Leo, where we
just posed the problem. Quite often when you form a collaboration you have
an idea of the solution and you put it together, but we had no idea why
this thing was working so well. So we plodded along and got various
insights along the way and I believe we figured it out (\cite{23}), at least to our satisfaction\ldots  but not to every
one's satisfaction: Leo never thought our explanation was the essential
reason. He thought our formal development was correct, but he didn't think
that was the reason that led to boosting's apparent spectacular
performance. But I was convinced that we had explained it. I think the
machine learners, the PAC learning people never thought so, I don't think
they completely understood the way we were looking at it.

\textbf{NF:} And since then you've had an extremely productive
collaboration with Rob and Trevor.

\textbf{JF:} Yes. We did that in the late 1990s, then later in the
mid-2000s I was asked to be an outside referee for a Ph.D. oral exam in the
Netherlands. A student of Jacq Meulman's, Anita vander Kooij, was
presenting her thesis and she had an idea. By this time the LASSO which Rob
Tibshirani had proposed in the mid-1990s was really coming on strong; it
still is. $L_{1}$-regularized methods and the LASSO, in particular, were
really becoming popular. There was a cottage industry on developing fast
algorithms for doing it. Engineers had worked on this, machine learners had
worked on this, and there was a spectacular paper by Brad Efron and some
colleagues (\cite{8}). So this was
very active at the time.

Then Anita and Jacqueline had this really simple idea that a professional
on optimization would dismiss out of hand, namely, just doing it one at a
time. They were working on a computer program that involved optimal
transformations of the variables, and for this they were using the
back-fitting algorithm. Including regularization then turned out to be
simple. Lots of people had developed the idea of optimizing one at a time.
This is usually dismissed in optimization theory as not performing well\ldots which is correct unless the one-at-a-time solution can be obtained
very conveniently and rapidly: then it can become competitive. Werner and I
had explored this with our so-called back-fitting algorithm in projection
pursuit regression and to fit additive models as well. Anyway, their idea
was that you hold all of the coefficients fixed but one and then solve for
the optimal solution for that one. This can be done very fast. Then you
just cycle through them. They developed it independently, but it was not a
new idea: other people had developed it before, but it didn't seem to have
been taken very seriously.

So when I came back from the Netherlands I told Rob and Trevor about this
and they got excited and we started working on applying the idea to a wide
variety of constrained and regularized problems and continue to do so to
this day. Rob and Trevor and their students have come up with all kinds of
new regularization methods, how we can do things one at a time and make it
go very fast. We applied it to the LASSO and to the Elastic Net, which was
something that Trevor and a student, Hui Zou, had done in the mid-2000s
(\cite{41}). It's a continuum of regularization methods between
ridge regression and the LASSO. You dial in how much variable selection you
want. In ridge there's no variable selection, LASSO does moderate variable
selection, so we extended it to the Elastic Net. Jacqueline and Anita had
also extended it to the Elastic Net. Then we extended it to other GLMs,
logistic regression, binomial, Poisson, Cox proportional hazards model, and
put together a whole package called \textit{glmnet} that seems to be widely
used now. It allows you to do all these different regularized regressions
with the various different GLM likelihoods, and that work is still going.

I like writing the programs because they seem to run faster than other
people's. It is probably because of my impoverished youth when I worked on
computers that were nothing like the computers now and you really had to
write efficient programs. That skill seems to have remained with me.

\textbf{NF:} This collaboration with Rob and Trevor resulted in a
particularly important publication.

\textbf{JF:} Yes, our book (\cite{37}). That
turned out to be an unbelievable success and I helped with parts of it, but
it was mostly written by Rob and Trevor. It just hit the right niche at the
right time and I guess it is still selling very well, but you can download
a pdf version from the Web for free now.

\textbf{NF:} Just to pick up on the point you made about you doing
the programming, I remember you told me years ago that you hadn't solved
the problem until you'd written the code to demonstrate the technique.

\textbf{JF:} I don't have the requisite skills to do all the theory. The
only way I can see if it's a good idea is if I program it up and try it
out, test it in a wide variety of situations and see how well it works.

\textbf{NF:} Let's pick up some parallel activities that you'd been
engaged in, starting with MART.

\textbf{JF:} At the time of my second lengthy visit to Australia in
1998/1999, I was fascinated with the boosting idea. MART was a kind of a
spin-off from the work that I'd done with Rob and Trevor on trying to
understand how boosting works. I got a few ideas for how to extend
boosting. Boosting was originally developed as a binary classification
problem and while I was visiting CSIRO in Sydney I wanted to extend it to
regression and to other kinds of loss functions, so I developed this notion
of gradient boosting which evolved into what I called MART, Multiple
Additive Regression Trees. I wrote that program (also called MART) and
developed those ideas. That was my Rietz Lecture I believe, which was
published in the \textit{Annals} (\cite{18}), an unusual paper for
them to publish.

I still wanted to understand more about why boosting was working. One of
the ideas that I had developed with the gradient boosting was the idea---again
a sort of a patience idea---that you can think of boosting as just
ordinary stepwise or stage-wise regression. You fit a model, say, a tree
(most people use trees), you take the residuals and then you fit a model to
the residuals. You take the residuals from the sum of those two trees and
build another model based on those residuals. Now that's very greedy; every
time you're trying to explain as much about the current residuals as you
can with the next model. I came up with an idea (again back to patient rule
induction!) that when one finds the tree that best fits the residuals, only
add a little bit of that tree, in other words, shrink its contribution. So
you multiply that tree by a small number like 0.1 or 0.01, before it's
added to the model. That turned out to really improve the performance. So I
wanted to understand why it was improving the performance and try to
understand more about gradient boosting.

%f4 #&#
\begin{figure*}

\includegraphics{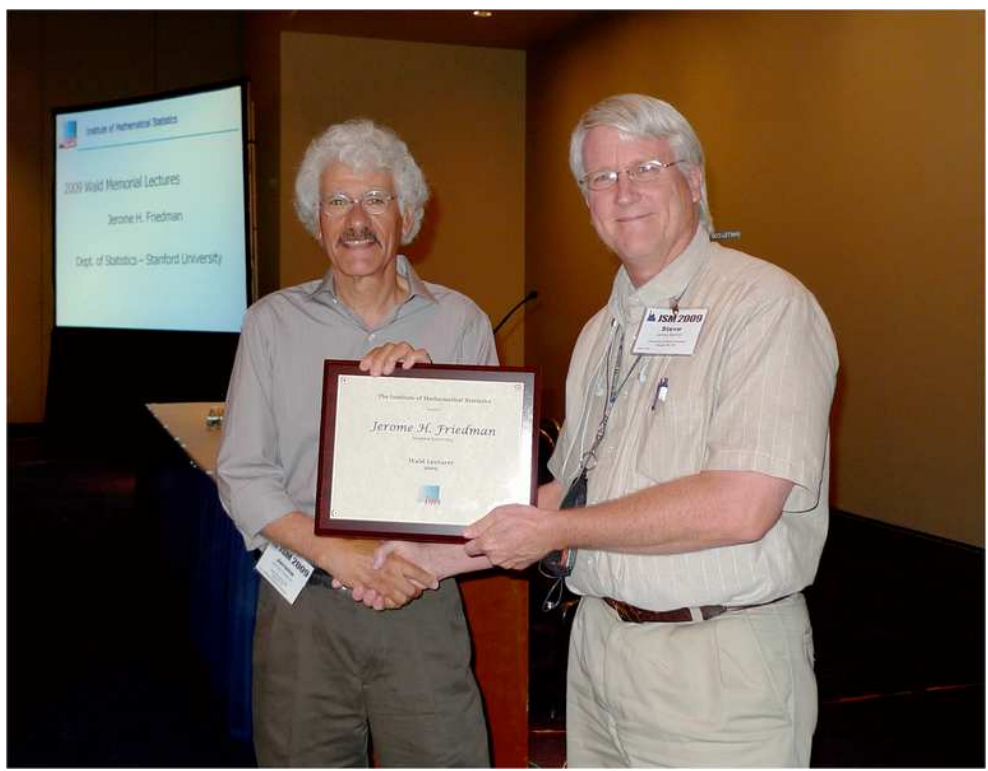}

\caption{Steve Marron presenting Jerry with his award for delivering the
Wald Lectures, Joint Statistical Meetings 2009. Photograph: Tati Howell.}%\label{fig4}
\end{figure*}

This was the time when Bogdan Popescu was my student. He showed that the
shrinkage only affected the variance and not the bias. I thought this was a
very important clue. Then, along with other people, we found that what this
was doing was a kind of LASSO. If you didn't do the shrinkage, then you
were doing something like stepwise or stage-wise regression. It produced
solutions that would be very similar to the LASSO and if you followed that
strategy in a linear regression, it produced solution paths very close to
the LASSO. In the beginning we thought they might be identical because we
ran a few examples and they produced identical paths. It turns out that it
will only produce identical paths in two dimensions or if the LASSO paths
are monotone functions of the regularization parameter.

Saharon Rosset did nice work in this area, as did others. There was part of
an issue in the \textit{Annals} devoted to boosting [\textit{Annals of
Statistics} \textbf{32}(1), 2004]. There were several very heavyweight
theoretical papers, very fine papers, showing that connection, showing that
boosting was consistent provided you regularized in this way.

\textbf{NF:} Jerry, you've had long-term enthusiasm for acronyms.
What do ISLE and RuleFit stand for?

\textbf{JF:} ISLE stands for Importance Sample Learning Ensembles. Again,
throughout this time I was interested in why the ensemble learning approach
was so effective and ISLE was a different way of looking at ensemble
methods. The idea was that you define a class of functions, and pick
functions from that class. The first thing that occurred to me was that
with boosting and bagging and other ensemble methods you just kept adding
trees. There were some people who thought, \textit{Okay, if you have an
ensemble, how do you figure out what is the optimal way to weight each
tree?} I thought that was a very simple problem: if I want to have a
function that's linear in a set of things, I know how to find the
coefficients, that's called regression. At this time, Leo was doing random
forests and a lot of people were doing boosting. I suggested that once you
get the ensemble, you just do a regularized regression to get the weights
of each of the trees or whatever they may be. Each element of the ensemble
in machine learning literature is called a base learner, or a weak learner,
because generally no one of them by themselves is very good, but the
ensemble of them is very good. That was one of the things that we
understood about why boosting worked. One of the reasons why boosting was
so surprising was that in machine learning literature they had a notion of
weak learners and strong learners: a weak learner is one that has low
learning capacity and a strong one has high. There was a lot of impressive
theoretical work by Rob Schapire, who was one of the co-inventors of the
original successful boosting algorithm. He showed that with this boosting
technique you could take a weak learner and turn it into a strong learner,
as long as the weak learner could achieve an error rate of $\varepsilon$
above 50\%. This was very lovely work.

But when you deal with it from this linear regression perspective it
doesn't seem so surprising. We've encountered many problems where just one
variable alone can't do much, but a number of variables fitted together in
a regression can do very well. From my statistical perspective that's
what's happening. I thought you could do this with a lot of different
things. If you have a class of functions, you pick functions from this
class and then you do a linear fit. Then the question is: how do you pick
the functions from the class? If you just randomly pick them, nearly all of
the functions will have no explanatory power as will their ensemble. If you
pick them to all be very strong, then their outputs are all highly
correlated and you are not gaining anything from the ensemble. The ensemble
will give the same predictions as any one of one of them.

So you have a trade-off that Leo had discussed a lot. You don't want your
learners in the ensemble to be highly correlated in their predictions, but
you do want them to have some predictive strength. That's a trade-off. This
was well known before the post-fitting idea. The LASSO and other
regularization methods are natural for the post-fitting because they could
be applied even when the size of the ensemble is much larger that the
number of observations. So you needed fast algorithms for the LASSO and
other regularized regressions that were being developed around that time.
It was a convergence of things.

RuleFit was an ensemble method totally motivated by this concept. The main
difference was that instead of fitting an ensemble with boosted trees and
then doing the post regression, you would take the trees, decompose them
into rules, forget the trees the rules came from, and use them as a batch
of ``variables'' in a linear fit.

Leo made a remark once, maybe in the mid-2000s shortly before his death,
that the real challenge in machine learning is not better algorithms,
grinding out a tiny bit more predictive accuracy. Our very best learning
machines tend to be black-box models---neural networks, support vector
machines, ensembles of decision trees---and they have very little if any
interpretive value. They may predict very well, but there is no way you can
tell your client why or how it is making a prediction, why it made that
prediction rather than another one. He thought that the real challenge was
interpretability and he had put some interpretational tools into his random
forest, namely, the relative importance of the predictor variables and some
other things. I wanted to see if there was some way to do interpretability
and the idea was that if you have an ensemble method, it's basically a
linear model and linear models are very interpretable as long as you can
interpret the constituents, the actual terms in the model. Trees you can
interpret, but I thought that it's easier to interpret rules. A tree
produces a rule derived from the path from the root to a terminal node:
that's why it's so interpretable. It can tell you exactly what variables
are used to make the prediction and how it used them, which is why trees
are so popular. Rule-based learning has also been a real staple in machine
learning throughout its history.

So I thought of breaking up the tree into its rules, putting the rules
together in a big pot and then doing a LASSO linear regression on the
rules. The hope was that since the rules aren't very complicated and are
easy to interpret, you could make much more interpretable models.

That in and of itself was only partially successful. But along the way I
developed ways for assessing the importance of the variables for individual
ensemble predictions. Another thing that I did in that work was to develop
some techniques for detecting interaction effects, seeing what variables
were interacting, exploring interaction patterns of the variables.

So that was RuleFit. I haven't done much beyond that in developing general
learning machines like MARS and MART, etc. RuleFit is my last one
so far.

\textbf{NF:} I dare say there will be more to come. You've been in
the Stanford department of statistics now for over thirty years. How have
you found it as an environment for a statistical scientist?

\textbf{JF:} Unbelievably great, I can't think of a place I'd rather be. My
greatest joy is to have an office in the hall with so many bright and
famous people. My nearest office neighbors are Brad Efron, Percy Diaconis
and Wing Wong, along with all the other fantastic people down the hall.
It's such a stimulating environment. Everyone is so sharp, so smart, so
inventive and original. You take it for granted after a while, but when you
visit other places you find it's not like that everywhere. I~consider it
great good fortune that I was able to join that Department and I thank them
for accepting me, because I was a kind of an odd appointment at the time.

\textbf{NF:} I am sure you look like a mainstream appointment right
now. Do you feel that the Department has, to some extent, progressed
towards you?

\textbf{JF:} Okay, maybe a little bit, yes.

%s6 #&#
\section{Current Interests}

\textbf{NF:} What are your current interests?

\textbf{JF:} There's the whole regularization idea which I still think is
fascinating. There are some leftover questions that current research has
not yet answered. I'd like to think more about that area. Another area is
improving decision trees. Trees have emerged as being very important
largely, in my view, because of the ensemble methods. Trees have very nice
robustness properties. They can be built quickly, they are invariant to
monotone transformations of the predictors, they are immune to outliers in
the predictors, they have elegant ways of handling missing values and of
incorporating both numeric and categorical variables. They are a very nice
type of learning machine: you just pour the data in, and you don't have to
massage the data too much prior to that.

They have several Achilles' heels, one of which was of course accuracy, but
I think that's been solved by the ensemble methods, which carry over all
these advantages while dramatically improving their accuracy: not just by
10\% or 20\%, but sometimes by factors of 3 or 4. I think boosting is one
of the key ideas of machine learning. It has really advanced both theory
and practice.

Another Achilles' heel of trees is categorical variables with a very large
number of levels. Back when we were doing CART, a typical categorical
variable might have 6 levels. Now it's routine to have hundreds or
thousands of levels. That destroys trees because there is no order
relation. The number of possible splits grows exponentially with the number
of levels. Optimizing over all these possibilities can lead to severe
over-fitting. In situations where there's a substantial amount of noise,
this can lead to spurious splits that mask the truly important ones.

So that's left over and it is one of those things that I mentioned I keep
in the back of my mind and every so often try to think about again, which
is what I'm doing now with this one.

Another thing I have been thinking about recently is the issue that many of
the problems arising with data that is seen now, especially commercial
data, tend to be binary classification problems. In my industrial
consulting I see much more classification than regression. This is
surprising because historically most statistics research has centered
around regression. Classification was something of a back issue in
statistics. In machine learning, classification has always been the main
focus. In fact, they refer to regression as classification with a
continuous class label.

A lot of the data is highly unbalanced, you may have millions of
observations but one class has very few. In engineering and machine
learning they tend to label the class as $+1$ and $-1$. Usually there is a
very small fraction of positives, like in fraud detection, for example,
where you have a data base with a huge amount of data, but the number of
instances of fraud is a small fraction of the data---at least you hope
that's the case! It's certainly true in e-commerce, where the rate of
clicking an ad on a page is around 1\% and then the conversion rate (which
means you click the ad and then and go buy something) is two orders of
magnitude lower than that. So the issue is how to deal with data like that
and there are rules of thumb that say if you have, say, a hundred positive
examples in a million negatives, you don't use all the million, you
randomly sample them. So then the question is: \textit{What's the strategy
and how many do you need?} And there's another rule of thumb that says if
you have 5 times as many negatives as positives, that's all you really
need. I doubt that's true in general, but I'd like to be more precise about
it because it's of huge practical importance: if you have millions of
observations which you can randomly sample down to a thousand or a few
thousand, that totally changes the dynamic of how you do your analysis. So
that's another thing I'm thinking about. Trevor Hastie and a student, Will
Fithian, recently did some nice work (\cite{10}) in this
area in the context of logistic regression.

Another area of current interest is loss functions. A machine learning
procedure is specified by a loss function on the outcome and a
regularization function on the model parameters. Defining appropriate
regularization functions and their corresponding estimators for different
problems is currently a hot topic for research in machine learning and
statistics. There is an avalanche of papers on the subject. There seems to
be less interest in finding appropriate loss functions for different
problems. The loss function $L(y, F)$ specifies the loss or cost
when the true value is $y$ and the model predicts $F$. I have found in my
consulting work that being able to customize the loss function for the
problem at hand can often lead to big performance gains. Most applications
simply use the defaults of squared-error loss for regression and Bernoulli
log-likelihood on the logistic scale for classification. I'd like to
investigate broader classes of loss functions appropriate for certain kinds
of specialized problems that go beyond the ones usually used in glms.

I find I spend a lot of my time on my programs. I~put most of my programs
on the Web and people can download them and use them, and they report bugs
back and I feel obligated to try to fix them. As you go along in your
career and you've done more and more things, you have to spend more and
more of your time back-caring---feeding those things---as well as
moving forward. I've had a long career now and spend a non-negligible
amount of my time just maintaining past stuff.

\textbf{NF:} {It's like entropy, isn't it, always increasing. The
list of errata never shrinks}.

\textbf{JF:} Yes. Then people have questions, they don't understand things,
or people use the algorithms in ways that you never dreamt they might be
used.

Something else I've just thought about. In the mid-1990s I worked a lot on
trying to incorporate regularization with nonconvex penalties. I spent a
fair amount of time on a technique which is somewhat similar to the
boosting technique but in the linear regression context. The LASSO imposes
moderate sparsity as opposed to an $L_{0}$-penalty (all subsets
regression) which induces the sparsest solutions. So I did a lot of work
spanning the gap between all subsets---which is very aggressive variable
selection and which often doesn't work, especially in low-signal settings---and the LASSO, which is moderately aggressive in selecting the
variables. That involves nonconvex penalties. The LASSO is the sparsest
inducing convex penalty. Of course with convex penalties, as long as you
have a convex loss function, then you have a convex optimization which is a
lot nicer than nonconvex optimization when you have multiple local minima
and other problems. So I did spend a lot of time working on boosting
techniques applied to linear regression with nonconvex penalties.

\textbf{NF:} {Statisticians around the world have been using your
techniques for a long time now, there's a company that exists simply to
sell your software and you generated that industry. Also your ideas and
methods were used by Yahoo!}.

\textbf{JF:} Yes. They used the commercial analog of MART as a big part of
their search engine. I don't know exactly what they use now, maybe the
Microsoft search engine. But for a long time, MART was an integral part of
the Yahoo! search engine.

%s7 #&#
\section{Life Outside Statistics}

\textbf{NF:} {Let's actually leave Statistics briefly, because you
do have a life outside Statistics}.

%f5 #&#
\begin{figure*}

\includegraphics{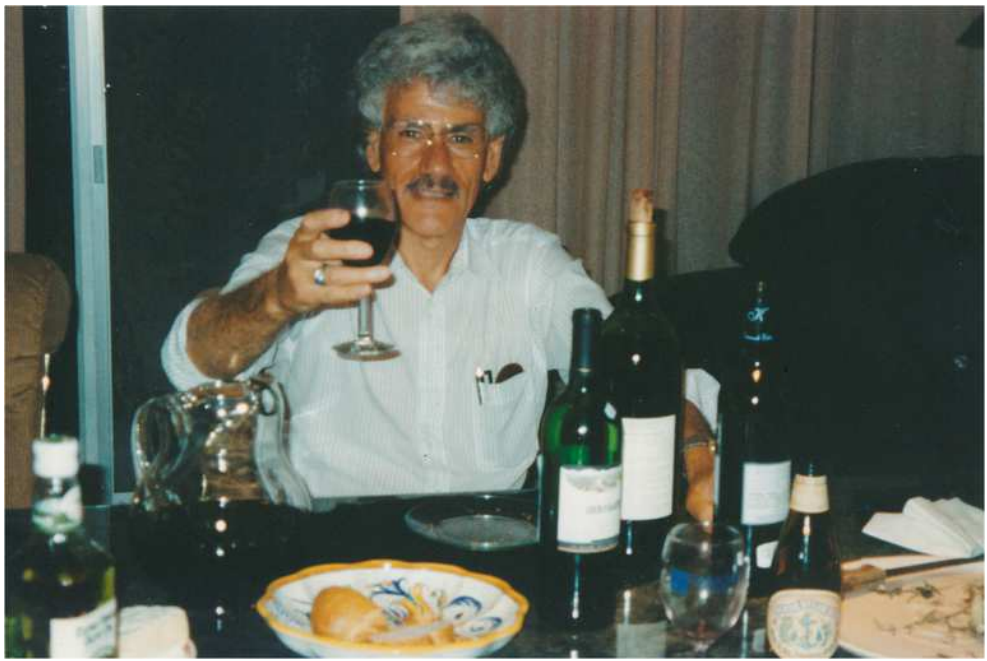}

\caption{A fine meal at home, 1997 Photograph: NIF.}%\label{fig5}
\end{figure*}

\textbf{JF:} Well, somewhat. (See the anecdote \textit{Life outside
Statistics} in \cite*{9}.)

\textbf{NF:} {And then there's been your long-time interest in
gambling and computers}.

\textbf{JF:} Yes, that started when I was a graduate student. (See the
anecdote \textit{Statistics, computers and gambling} in \cite*{9}.)

%s8 #&#
\section{Back to the Future}

\textbf{NF:} Finally, let's step back, or maybe move to a greater
height in this conversation, in the sense of taking a perspective on
statistics at certain times. There have been at least two occasions
(Friedman, \citeyear{15}, \citeyear{19}) when you have committed your thoughts to
print about ``Where are we now with statistics and computing?''
Let's go back to 1987 when there was a symposium on ``Statistics in
Science, Industry and Public Policy.'' You were invited to present
a paper on ``Modern Statistics and the Computer Revolution.''

\textbf{JF:} It was an assignment I couldn't refuse because it was from the
person in charge of statistics funding at NSF. I had an NSF grant at the
time, so I had to go back and give a talk about what I thought about the
future of statistics and how computing might affect statistics in the
future.

\textbf{NF:} In this paper you talked about automatic data
acquisition, some of its benefits and also some of the issues that it
raised. Early in the paper you said that ``What separates Statistics to a
large degree from the information sciences is that we seek to understand
the limits of the validity of the inference.'' Do you think that
separation is still the case in, say, machine learning areas?

\textbf{JF:} Not as much as it was, but I would say so. The huge
contribution of statistics to data analysis is inference, what you are
getting out of the data, or learning from the data. How much of it is
really valid. That has been the main thrust of statistics. It has become
less of a thrust only because data sets have gotten larger and so the
sampling variation has become less of a problem, but it's still there in a
big way. Originally, I think there were people in neural networks and
machine learning who weren't very concerned about that at all, whatever
they found they assumed was reality. And to be fair, at least in machine
learning, that was because they were dealing with pattern
recognition---problems where the inherent noise was not large, where the Bayes error rate
in a classification problem was really very close to zero if not zero. The
particular classifier that attained that error rate was complicated and
hard to get at. So I don't think that inference was as big a problem in
those kinds of things. Statisticians originally came from other areas where
the data sets were small and signal to noise was very low. In those
settings inference is a very important part of the learning procedure.

\textbf{NF:} But the computer scientists and the machine learners
haven't stayed in their little box, they started playing with other
problems.

\textbf{JF:} Oh yes, Bayes-type ideas are now spread\linebreak[4] throughout machine
learning, computer science and engineering, for example. Inference is
there, although it's perhaps not given quite the high priority that we
statisticians give it.

\textbf{NF:} You commented that most of the methods being used in
Statistics in 1986 were actually developed before 1950, but that the
computer was liberating us from these mathematical bindings such as
closed-form solutions and unverifiable assumptions. I particularly like
your closing comment that ``The cost of computation is ever decreasing but
the price we pay for incorrect assumptions is still staying the
same.'' Would you care to amend that statement now?

\textbf{JF:} No, I think it's the same, we have to make fewer and fewer
unverifiable assumptions these days. The sample reuse techniques like
cross-validation and the bootstrap have really freed us up; they have
really helped the kind of thing I do a lot. Quite often when you come up
with a new complicated procedure and someone will say, ``How do you do the
inference? How do you put error bars in?'' or something like that, you just
reply, ``Well you can bootstrap it.'' So that was a giant contribution to
statistics. But in the area that I work in it is especially valuable.

\textbf{NF:} Moving on 12 years, you had another opportunity to
take a helicopter view at the ISI meeting in Helsinki, where there was a
session on ``Critical Issues for Statistics in the Next Two
Decades.'' You presented a paper on ``The Role of Statistics in
the Data Revolution?,'' and I note the question mark at the end of
that statement! In the summary you said,
``The nature of data is rapidly changing. Data sets are becoming
increasingly large and complex. Modern methodologies for analysing these
new types of data are emerging from the fields of data base management,
artificial intelligence, machine learning, pattern recognition, and data
visualization. So far, statistics as a field has played a minor role. This
paper explores some of the reasons for this and why statisticians should
have an interest\ldots''
and so on. What I'm interested in is: How have things changed since
then, what needs to be done, and what's blocking this change?

\textbf{JF:} Oh, I think it's changing quite a bit. Perhaps I have a
nonrepresentative view being at Stanford, but I think that statistics is
definitely moving forward in those areas. Statistical research in data
analysis is definitely overlapping more with machine learning and pattern
recognition. As I pointed out in the 1987 paper, and as I say whenever I am
asked about the future of statistics, you can't answer that question, you
have to ask: \textit{What is the future of data?} Statistics and all of the
data sciences will respond to whatever data is present. No one could have
anticipated gene expression arrays in the late 1980s. Now statisticians
have adapted to that and the whole bioinformatics revolution as well,
making huge contributions to those areas.

\textbf{NF:} In particular, in the 1999 paper, reflecting on the
relationship between statistics and data mining, you said that ``From the
perspective of statistical data analysis, however, one can ask whether data
mining methodology is an intellectual discipline. So far the answer is: Not
yet\ldots '' Has the answer changed or has the question become
irrelevant?

\textbf{JF:} That's a good question. I would say it's relevant and
changing; I'm not sure that it has totally changed yet. I think you need
people who can come up with several ways of looking at data but who perhaps
don't have the requisite skills to understand at a basic level what's
happening. And you need people who are very skilled at taking a methodology
and a situation and then deriving the properties of the method in that
situation. I think the attitude in the data mining community is: ``If it
works, great! We'll try things and we'll find out the things that work.'' I
think that's a perfectly reasonable way to proceed. Some people like to
proceed from basic principles: let's first understand the basic principles,
and from there develop the right things to do, or good things to do. The
other is an \textit{ad hoc} approach---just think hard about the problem,
try to figure it out---which is the way Tukey did it way back when---and try to come up with something that works well. That approach is fraught
with danger, of course: not everyone is as smart as Tukey. As people
develop techniques, they advertise and convince people they are really very
good when they are not, so one has to be careful. But generally, if there's
a methodology like PLS, Support Vector Machines, boosting or more general
ensemble methods that seems to repeatedly work very well, there's probably
a good statistical reason, even if in the beginning it was not known. The
understanding, the underlying principles of why they work well, came later
on.

\textbf{NF:} Later in that paper you said, ``Perhaps more than any
other time in the past statistics is at a crossroads; we can decide to
accommodate or resist change.'' {Have we accommodated, are we still
resisting change, how do you situate statistics now in the information
sciences?}

\textbf{JF:} I think statistics is accommodating change, not as fast as I
would like, faster than some other people would like, but certainly
adapting to change. Ultimately it is data that's driving statistics as well
as the other information sciences. But I think statistics today is much
more responsive. When new forms of data come out there are statisticians
who immediately see the opportunity, as well as engineers and other
people.

\textbf{NF:} Then in a sense I think you have answered your
concluding remark in this paper, which was: ``Over the years this
discussion has been driven mainly by two leading visionaries of our field.
John Tukey in his 1962 Annals of Mathematical Statistics paper (\cite{38}) and
Leo Breiman at the 1977 Dallas conference. Over twenty years have
passed since that conference. We again have the opportunity to re-examine
our place among the information sciences.'' So you feel that we are
sitting rather more comfortably in there than we did?

\textbf{JF:} Again, being at a wonderful place like Stanford, I think so,
yes. I think we are doing it right. We haven't abandoned our tradition of
formal inference, which is very good because that's something that the
other information sciences don't do nearly as well as we do. There are
isolated incidences of people in those other areas who do it very well, but
it's not the priority that it is in statistics. That being one of our
priorities really helps a lot because you must understand the limits of
inference at some point. I think in the early stages we were trying things
out, seeing what would work and using our intuition and I think the
insights are beginning to come. These days, if you look at the work being
done in bioinformatics, in computer science and in statistics, there's a
huge overlap where there wasn't before\ldots  in attitude as well as the actual
work, the problems we're trying to solve.

\textbf{NF:} Well, we still hold true to the guiding standard of
understanding and managing variability.

%f6 #&#
\begin{figure*}

\includegraphics{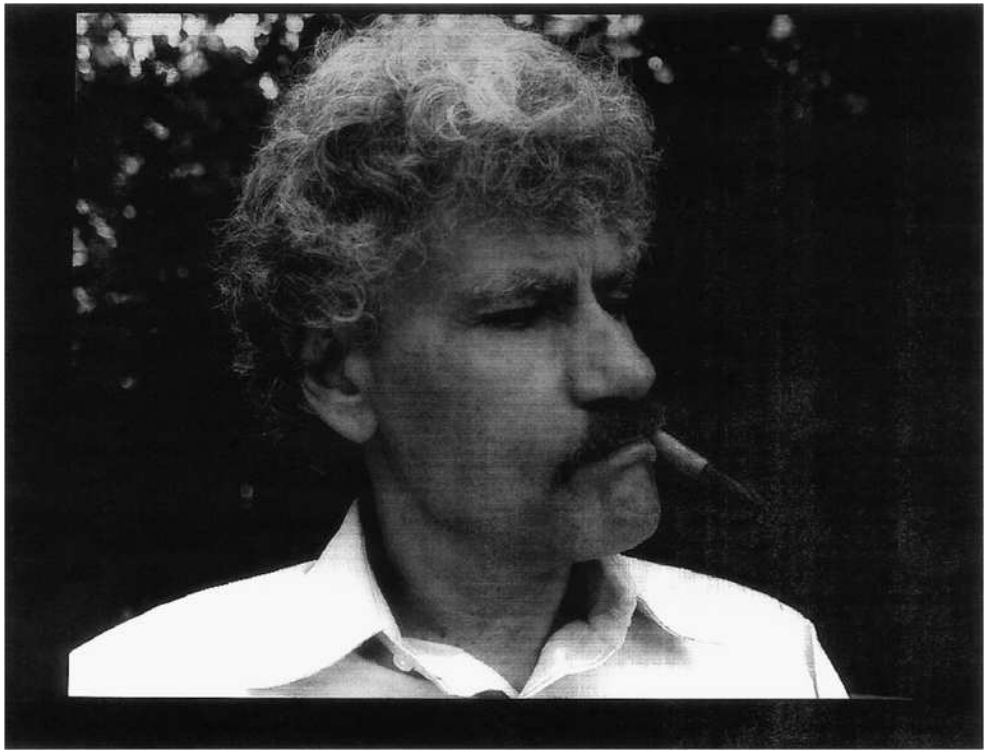}

\caption{A fine cigar. Photograph: Ildiko Frank.}%\label{fig6}
\end{figure*}

\textbf{JF:} That's right, and I think we pay more attention to that than
other fields do and I think that's good. In the past, perhaps we may have
paid too it much attention. Well, not \textit{too} much attention because
statistics was working on methodology for a certain kind of data---small
data sets, high noise, where inference was everything, are you seeing a
signal or not? This is the essence of hypothesis testing. Not \textit{How
big is the signal and what are its properties?}, just \textit{Can we say
whether there is one or not?} With those small data sets and high noise
often that was the only thing you could ask. Hypothesis testing was a huge
intellectual triumph. But now with larger data sets and better
signal-to-noise ratios, we can start asking more detailed questions:
\textit{What is the nature of the signal? What variables are participating
in the prediction problem? How are they participating? How are they working
together to produce the result?}

\textbf{NF:} You have certainly changed the way a lot of people
think about Statistics and you believe you are doing Statistics and have
done consistently. If Colin Mallows were in our presence now, do you think
he would be describing what you do as Statistics?

\textbf{JF:} I guess you'd have to ask him. Perhaps. I have always believed
that---perhaps erroneously!---but I always believed that I was doing
statistics. You know what they say: a rose by any other name would smell as
sweet. I think there is less of a need to categorize things. Who cares
about the name of what you're doing as long as it's interesting and
potentially useful. The categories seem to be all blurred now and that's
all to the good.

\textbf{NF:} Okay, well by a miracle of modern science we have
sitting beside us a reincarnation of Jerry Friedman, except he's only
twenty years old, and he's wondering what to do at college. What are you
going to recommend?

\textbf{JF:} What I always recommend whenever I'm asked: ``What should I
study, what should I do?'' I always say, ``Study and follow what you are
most passionately interested in. Don't worry about what skills are going to
be marketable in ten years because that will all change.'' If you go to
school to learn a skill that you don't like because you think it is going
to be especially marketable when you get out 5 or 6 or 8 years from now,
that could change. You've suffered through all of that and you end up
without marketable skills after all. At least if you study something you're
really enjoying or are passionate about, you've had all that fun. If you're
lucky like I was and it turns out that your skill evolves into being
marketable, then so much the better. Follow your passion.

\textbf{NF:} You think statistics might easily be one of those?

\textbf{JF:} Oh, I agree with Hal Varian (Chief Economist, Google), who
made that statement, that statistics is going to be the glamor field of the
future for some time (``I keep saying the sexy job in the next ten years
will be statisticians.'' \cite*{39}). People think I'm joking, but who
would've guessed that computer engineers would've been the sexy job of the
1990s? The data revolution---using data to answer questions and solve
problems---has really emerged. Not so long ago when, say, you were at a
factory or at some kind of production line and yield was going down, what
did you do about it? Well, you called on the supervisors and experts, you
got into a room and you tried to figure why yield might be going down. It
didn't often occur to people to collect data. Now everybody collects data.
Almost every production line and factory is heavily instrumented at every
point and data is being collected. In fact, I think maybe it may come to
the point where people ask too much of data; data can't answer every
question.

\subsection*{CODA}
%\noindent\rule{\columnwidth}{0.5pt}

\textbf{NF:} I was pondering how to title this conversation and I
did have in mind something like ``Jerry's search for pattern,'' but then it
occurred to me that a pattern is only a pattern\ldots

\textbf{JF:} \ldots but a good cigar is a Smoke. I agree.

\textbf{NF:} Somebody once said something along those lines.

\textbf{JF:} Yes, it was Kipling of course (e.g., \cite*{31}). I
started smoking cigars on and off when I was young, in high school and just
out of high school. I~worked for the Forestry Service fighting forest fires
and surveying timber access roads. Where I lived most of the countryside
was national forest and so that was a traditional job to do. At one of the
camps the only facilities were out-houses and they smelled very, very bad.
It was a real ordeal to use them, especially if you had to stay longer than
ten or twenty seconds. The only way that I could stand to do it was to
light up a really foul-smelling cigar, and smoke it while I was in there.
That's why I started smoking cigars. I smoke better cigars now.

\textbf{NF:} So do you feel we should stop talking about patterns
right now and adjourn\ldots?

\textbf{JF:} It wouldn't be a bad idea.

\textbf{NF:} Well then, many thanks, Jerry, for this glimpse of a
fascinating scientific odyssey. I feel as if I've been slip-streaming Slim
Pickens, riding a rocket down the years in which statistics and computing
have become inextricably intertwined, except you've been sitting on the
nose-cone and pointing the rocket, which Slim Pickens didn't quite have the
ability to do. May you ride for a long time to come.

\textbf{JF:} Well, thank you very much, Nick, I really appreciated it.

\section*{Acknowledgments}
The author thanks Rudy Beran and Bill van Zwet for
valuable critical comment on a draft of this article, the Editor and an
Associate Editor for their helpful feedback, and Jerry for his hospitality
during the interview and patience during the preparation of the
article. The work was supported in part by ValueMetrics Australia.

\begin{supplement}%[id=suppA]
\stitle{Supplement to ``A conversation with Jerry Friedman''}
\slink[doi]{10.1214/14-STS509SUPP} %[doi,text={...}] - jei reikia suskaldyti doi
\sdatatype{.pdf}
\sfilename{sts509\_supp.pdf}
\sdescription{The supplementary materials associated with this article comprise a number
of anecdotes, plus an example of one way in which John Tukey communicated
his research ideas to Jerry in the course of their collaboration. They are
available from \citet{9}.}
\end{supplement}

\iffalse
%s9 #&#
\section{Supplementary materials}

The supplementary materials associated with this article comprise a number
of anecdotes, plus an example of one way in which John Tukey communicated
his research ideas to Jerry in the course of their collaboration. They are
available from \citet{9}.

\fi
%\begin{appendix}
%\section{}
%\end{appendix}


% imsref loaded by imikolaityte, 2015-01-20 15:06:51

\end{document}